\documentclass[fleqn,usenatbib]{mnras}

\usepackage[T1]{fontenc}
\usepackage{ae,aecompl}
\usepackage{graphicx}	
\usepackage{amsmath}	
\usepackage{amssymb}	
\usepackage{amssymb}
\usepackage{placeins}

\newcommand\tempbf[1]{#1}
\newcommand\tempbfnew[1]{#1}

\title[The \texttt{THOR+HELIOS} general circulation model]{The \texttt{THOR+HELIOS} general circulation model: multi-wavelength radiative transfer with accurate scattering by clouds/hazes}

\author[Deitrick et al.]{Russell Deitrick$^1$\thanks{E-mail: russell.deitrick@unibe.ch (RD)}, Kevin Heng$^{1,2}$, Urs Schroffenegger$^1$,  Daniel Kitzmann$^1$, \newauthor
Simon L. Grimm$^1$, Matej Malik$^3$,  Jo\~{a}o M. Mendon\c{c}a$^4$, and Brett M. Morris$^{1,5}$
\\
$^{1}$University of Bern, Center for Space and Habitability, Gesellschaftsstrasse 6, CH-3012, Bern, Switzerland\\
$^{2}$University of Warwick, Department of Physics, Astronomy and Astrophysics Group, Coventry CV4 7AL, UK\\
$^{3}$University of Maryland, Department of Astronomy, 4296 Stadium Drive, College Park, MD 20742, U.S.A.\\
$^{4}$Technical University of Denmark, National Space Institute, Astrophysics and Atmospheric Physics, \\ Elektrovej, DK-2800, Kgs. Lyngby, Denmark\\
$^{5}$University of Bern, Physics Institute, Division of Space Research \& Planetary Sciences, \\ Sidlerstrasse 5, CH-3012, Bern, Switzerland}

\date{}

\pubyear{2020}

\begin{document}
\label{firstpage}
\pagerange{\pageref{firstpage}--\pageref{lastpage}}
\maketitle

\begin{abstract}
General circulation models (GCMs) provide context for interpreting multi-wavelength, multi-phase data of the atmospheres of tidally locked exoplanets.  In the current study, the non-hydrostatic \texttt{THOR} GCM is coupled with the \texttt{HELIOS} radiative transfer solver for the first time, supported by an equilibrium chemistry solver (\texttt{FastChem}), opacity calculator (\texttt{HELIOS-K}) and Mie scattering code (\texttt{LX-MIE}).  To accurately treat the scattering of radiation by medium-sized to large aerosols/condensates, improved two-stream radiative transfer is implemented within a GCM for the first time.  Multiple scattering is implemented using a Thomas algorithm formulation of the two-stream flux solutions, which decreases the computational time by about 2 orders of magnitude compared to the iterative method used in past versions of \texttt{HELIOS}.  As a case study, we present four GCMs of the hot Jupiter WASP-43b, where we compare the temperature, velocity, entropy, and streamfunction, as well as the synthetic spectra and phase curves, of runs using regular versus improved two-stream radiative transfer and isothermal versus non-isothermal layers.  While the global climate is qualitatively robust, the synthetic spectra and phase curves are sensitive to these details.  A \texttt{THOR+HELIOS} WASP-43b GCM (horizontal resolution of about 4 degrees on the sphere and with 40 radial points) with multi-wavelength radiative transfer (30 k-table bins) running for 3000 Earth days (864,000 time steps) takes about 19-26 days to complete depending on the type of GPU.
\end{abstract}


\section{Introduction}
\label{sect:intro}

\begin{table*}
\label{tab:GCMs}
\begin{center}
\caption{Selected list of GCMs used for hot Jupiters}
\begin{tabular}{lcccc}
\hline
\hline
Name of code & Dynamical equations & Multi-wavelength & Radiative transfer & Reference(s) \\
 & solved & radiative transfer? & method & \\
\hline
\hline
\texttt{SPARC/MITgcm} & Primitive & Yes & Two-stream source & \cite{showman09} \\
 & & & function method & \\
\texttt{IGCM} & Primitive & No$^\dagger$ & Two-stream & \cite{rm10,rm12} \\
\texttt{FMS} & Primitive & No$^\dagger$ & Two-stream & \cite{heng11a,heng11b} \\
--- & Non-hydrostatic Navier-Stokes & Yes$^\ddagger$ & Flux-limited diffusion & \cite{dd12} \\
--- & Non-hydrostatic Navier-Stokes & Yes & Two-stream & \cite{dd13} \\
\hline
\texttt{UM} & Non-hydrostatic Euler & Yes & Two-stream & \cite{mayne14a,mayne14b,mayne17} \\
 & & & (Edward-Slingo$^\star$ method) & \cite{amundsen16} \\
\hline
\texttt{THOR v1} & Non-hydrostatic Euler & No$^\dagger$ & Two-stream & \cite{mendonca16} \\
\texttt{THOR v2} & Non-hydrostatic Euler & No$^\dagger$ & Two-stream & \cite{deitrick20} \\
 \texttt{THOR+HELIOS} & Non-hydrostatic Euler & Yes & Improved two-stream & Current study \\
\hline
\hline
\end{tabular}\\
{\scriptsize $\dagger$: Used dual-band or ``double-gray" radiative transfer that requires the specification of two mean opacities (in the optical and infrared). \\ $\ddagger$: Stellar energy deposition is multi-wavelength in implementation, but radiative fluxes are computed using Rosseland mean opacities.\\ $\star$ \cite{Edwards1996} }
\end{center}
\end{table*}

\subsection{Providing context for interpreting multi-wavelength, multi-phase data}

Hot Jupiters are tidally locked, highly irradiated, hydrogen-dominated exoplanets \citep{burrows10,fortney10}. They are, of course, also three-dimensional (3-D) objects. Thus, to fully understand their atmospheres requires the procurement of emission and transmission spectra at different orbital phases or phase curves at different wavelengths (see \citealt{showman10,burrows14b,hs15,par15,zhang20,showman20} for reviews), as well as constraints on their variability (e.g., \citealt{agol10}).  With the \textit{Hubble Space Telescope} (HST), measuring multi-wavelength phase curves is restricted to hot Jupiters on short (<1 day) orbits.  To date, these include WASP-43b \citep{stevenson14}, WASP-103b \citep{kreidberg18} and WASP-18b \citep{arc19}.  With the upcoming \textit{James Webb Space Telescope} (JWST), the procurement of multi-wavelength phase curves---and even eclipse maps \citep{dewit12,majeau12}---of hot Jupiters is expected to become routine.

There is a rich body of literature on using one-dimensional radiative-convective models to interpret the spectra of hot Jupiters (e.g., \citealt{ss98,ss00,sudarsky00,sudarsky03,barman01,barman05,burrows03,burrows07a,burrows07b,burrows08a,burrows08b,fortney05,fortney08,tinetti07,sb10}); see \cite{burrows14a} for a review.  Without resorting to parametrisations of the three-dimensional dynamical and thermal structure, it is unclear how to self-consistently compute the dayside emission spectrum, nightside emission spectrum, transmission spectrum (associated with the terminator regions), multi-wavelength phase curves and temporal variability of a tidally locked exoplanet.  Some promising two-dimensional (2-D) or pseudo 2-D approaches have been implemented \citep{tremblin17,gandhi20}, but these generally still require 3-D models for tuning and/or validation. To this end, general circulation models (GCMs) are an essential tool for understanding the relationship between atmospheric dynamics, radiation, chemistry and the observational signatures of tidally locked exoplanets. GCMs also provide context for atmospheric retrieval techniques that use one-dimensional radiative transfer models (e.g., see \citealt{madhu18,bh20} for reviews). Further, 3-D information from GCMs is now being utilized directly in atmospheric retrievals \citep{flowers19,beltz21,wardenier21}. 

\subsection{Moving beyond Solar System-centric general circulation models}

GCMs were originally developed for the study of the climate of Earth (e.g., \citealt{wp05}).  There is a long and enduring legacy of Earth GCMs (e.g., \citealt{adcroft04,anderson04,fhz06,fhz07,os08}) and the study of the terrestrial climate as a heat engine \citep{po84}.  The need to benchmark dynamical cores (the code that solves the fluid equations) was recognized by \cite{hs94}.  Model hierarchies were proposed by \cite{held05} as an approach for attaining deeper understanding of the ingredients of GCMs and how they interact.

For Jupiter, GCMs were developed as part of a debate on whether the Jovian jet/wind structures are shallow (e.g., \citealt{cp96}) or deep (e.g., \citealt{kaspi09,sl09}); see \cite{vs05} for a review.  This debate was settled recently for Jupiter \citep{kaspi18} and also for Neptune and Uranus \citep{kaspi13}.  A lesson learned from Jupiter GCMs is that the primitive equations of meteorology (see Appendix \ref{append:review} for a review) are  suitable for hot Jupiters, as long as the model domain is relatively small compared to the radius \citep{mayne14b,mayne17,deitrick20}.  Other lessons are difficult to generalize as Jupiter is a fast rotator (with a Rossby number well below unity), whereas tidally locked hot Jupiters are slow rotators (with Rossby numbers on the order of unity), implying that their jet/wind structures are qualitatively distinct \citep{menou03}.  Furthermore, the energy budgets of hot Jovian atmospheres are dominated by stellar irradiation, rather than heating from the deep interior, implying that convection is suppressed and equator-to-pole circulation is present \citep{heng11b}.  The high ($\gtrsim 1000$ K) temperatures of hot Jovian atmospheres imply that the dominant opacity sources will be different from their Solar System counterpart.

The study of the atmospheric dynamics of hot Jupiters was pioneered by \cite{sg02} and \cite{gs02}.  The early works of \cite{cho03,cho08}, \cite{menou03} and \cite{mr09} treated only a shallow layer of hot Jovian atmospheres.  \cite{cs05}, \cite{showman08,showman09} and \cite{rm10} were the first to use GCMs to model the deep atmospheres of hot Jupiters.  \cite{showman09} was the first study to build multi-wavelength radiative transfer into a hot Jupiter GCM, an effort that was followed up by \cite{amundsen16}.  \cite{lewis10} and \cite{kataria13} used GCMs to study irradiated exoplanets on highly eccentric orbits, building on the work of \cite{ll08}.  \cite{heng11a} generalized the GCM benchmark tests of \cite{hs94} for tidally locked exoplanets, based on the work of \cite{mr09}, \cite{ms10} and \cite{rm10}.  \cite{heng11b} and \cite{rm12} introduced dual-band or double-gray radiative transfer into hot Jupiter GCMs.  \cite{dd10,dd12} and \cite{dd13} adapted a fully explicit, non-hydrostatic fluid dynamical solver, albeit with a truncated grid, to study hot Jovian flows and their observational signatures.  \cite{mayne14a,mayne14b,mayne17} adapted a sophisticated Earth weather/climate model with a non-hydrostatic solver and applied it to hot Jupiter atmospheres. Concurrently, \cite{mendonca16} introduced a flexible, non-hydrostatic GCM built from scratch (\texttt{THOR}, the model used in the present work). \cite{perna12} used a suite of GCMs to explore the effects of varying stellar irradiation.  \cite{ls13} demonstrated the insensitivity of hot Jupiter GCM outcomes to initial conditions.  \cite{par13} studied the interaction between atmospheric dynamics and condensates in hot Jupiter GCMs.  \cite{kataria15} compared GCM outputs of WASP-43b to emission spectra measured at different orbital phases, an effort that was followed up by \cite{mendonca18a}.  \cite{oreshenko16} investigated the effects of scattering by condensates in simplified GCMs of Kepler-7b.  \cite{ks16} and \cite{ks17} elucidated the mechanism underlying dayside-nightside heat redistribution, building on the work of \cite{sp10,sp11}.  \cite{drummond18} and \cite{mendonca18b} implemented a simplified disequilibrium chemistry scheme known as ``chemical relaxation" \citep{cs06} in hot Jovian GCMs, while \cite{steinrueck19} focused on the observational consequences of disequilibrium. \cite{drummond20} combined two-stream radiative transfer with chemical kinetics into a fully self-consistent hot Jupiter GCM. Meanwhile, several works have focused on flaws of GCM use for hot Jupiters \citep{thrastarson11,skinner21}. While the vast majority of models show prograde equatorial flow, some have suggested that retrograde flow may be possible \citep{mendonca20,carone20}. \cite{sainsbury20} have explored the role of potential temperature mixing in the atmosphere on the ``radius-inflation'' problem using GCM simulations. \cite{lee20} used a GCM to study a brown dwarf that straddles the line between hot Jupiters and stars, as it is highly irradiated by a white dwarf companion. Most recently, a wealth of studies have begun to examine the effects of condensates in hot Jupiter atmospheres, both with gray \citep{roman17,mendonca18a,roman19,roman21} and non-gray \citep{par16,lines18,lines19,par21} radiative-transfer. 

As summarised in Table \ref{tab:GCMs}, most of the GCM studies cited in the preceding paragraph use one of the following GCMs: \texttt{SPARC/MITgcm} \citep{showman09}, the \texttt{IGCM} \citep{mr09,rm10,rm12}, the \texttt{FMS} (e.g. \citealt{hs94,fhz06}), the U.K. Met Office \texttt{UM} \citep{mayne14a,mayne14b,mayne17,amundsen16} or \texttt{THOR} \citep{mendonca16,deitrick20}. The computer code of \cite{dd08}, \cite{dd10,dd12} and \cite{dd13} is unnamed.

The current study builds on this rich body of work on hot Jupiter GCMs by coupling the \texttt{THOR} GCM with the \texttt{HELIOS} radiative transfer solver \citep{malik17,malik19} for the first time, building on the pioneering work of \cite{showman09,amundsen16,mayne17} and others.

\subsection{Accurate scattering of radiation by medium-sized and large aerosols/condensates}
\label{subsect:accurate_scattering}

Hot Jupiters are believed to have cloudy/hazy atmospheres (e.g., \citealt{pont08,pont13,sing16,stevenson16}), which motivates the accurate treatment of scattering by aerosols/condensates.  The two-stream method of radiative transfer is used extensively in GCMs (Table 1), because of its speed and simplicity of implementation.  However, it suffers from a serious shortcoming: it over-estimates the backscattering of radiation caused by medium-sized and large aerosols \citep{kitzmann13,hk17}.  In one-dimensional climate models of early Mars, this artefact has been demonstrated to produce $\sim 50$--70 K of artificial warming by the scattering greenhouse effect \citep{kitzmann16}. While some GCMs (\texttt{SPARC/MITgcm}, \texttt{UM}) have implemented more accurate two-stream methods, the effects of different two-stream solutions on hot Jupiter GCMs remain under-explored.

\cite{hk17} and \cite{heng18} proposed an improved two-stream method of radiative transfer, which removes the artefact of too much backscattering by calibrating the ratio of Eddington coefficients to 32-stream discrete ordinates calculations.  Here, we refer to the \texttt{HELIOS} solution that uses the backscattering correction as ``improved two-stream'' and the \texttt{HELIOS} solution without the correction as ``regular two-stream''. In the current study, the improved two-stream method is implemented within a GCM.  By comparing the outputs from a pair of GCMs implementing the regular versus improved two-stream methods, we will quantify the error incurred when simulating hot Jupiters.  Multiple scattering of radiation is handled using a matrix formulation of the improved two-stream flux solutions, where a tridiagonal matrix is inverted using Thomas's algorithm. 

\subsection{Structure of paper}

\begin{figure*}
\begin{center}
\vspace{-0.1in}
\includegraphics[width=\textwidth]{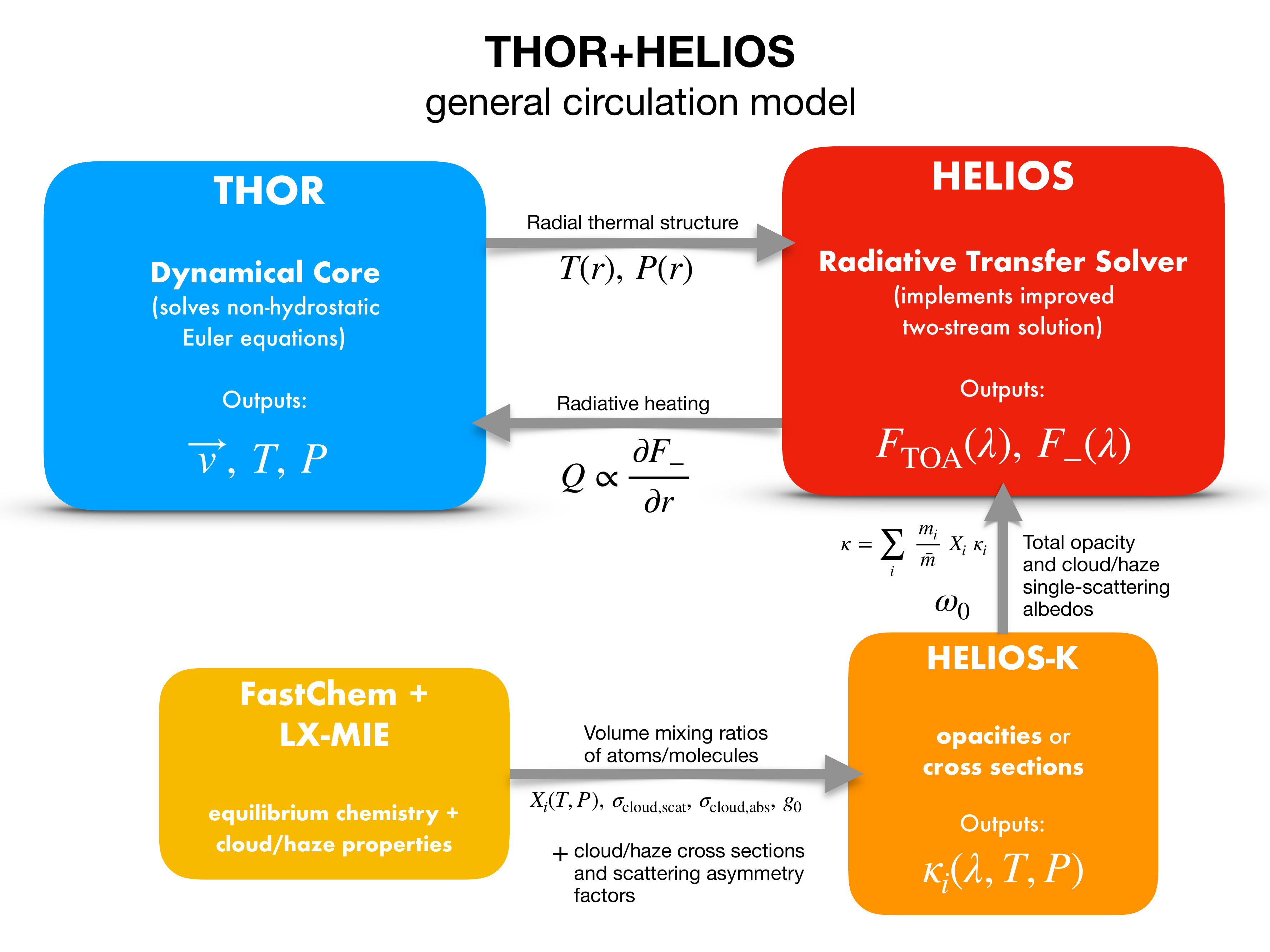}
\end{center}
\vspace{-0.2in}
\caption{Overview of the \texttt{THOR+HELIOS} general circulation model, which mainly consists of a dynamical core (\texttt{THOR}) and a radiative transfer solver (\texttt{HELIOS}) and is supported by an opacity calculator (\texttt{HELIOS-K}), equilibrium chemistry solver (\texttt{FastChem}) and Mie scattering code (\texttt{LX-MIE}).  Given the three-dimensional velocity ($\vec{v}$) and thermal structure ($P(r), ~T(r)$) provided by \texttt{THOR}, \texttt{HELIOS} performs radiative transfer and computes the radiative net fluxes ($F_-$) and the flux emerging from the top of the atmosphere (TOA; $F_{\rm TOA}$). The iteration between \texttt{THOR} and \texttt{HELIOS} solves for a general equilibrium between the three-dimensional atmospheric dynamics and radiative heating, which is more general than the radiative or radiative-convective equilibrium typically computed by one-dimensional radiative transfer codes.}
\vspace{-0.1in}
\label{fig:schematic}
\end{figure*}

The current study is the culmination of a decade of theoretical and computational developments published in more than a dozen papers (Table 2).  The foundation of these developments is the first version of \texttt{THOR} by \cite{mendonca16}.  As such, a substantial fraction of the current paper is devoted to first concisely reviewing these developments (for self-contained readability) and subsequently integrating them into a single entity.  Figure \ref{fig:schematic} provides an overview of how the different components of \texttt{THOR+HELIOS} operate together.  Section \ref{sect:methods} contains a detailed description of both previous and novel methodology. Section \ref{sect:benchmark} presents 1-D comparisons between the new code and the standalone version of \texttt{HELIOS} and tests of the spectral convergence. Section \ref{sect:results} presents an illustrative set of four WASP-43b GCMs computed using \texttt{THOR+HELIOS}.  Section \ref{sect:discussion} provides a summary of the key developments and findings, as well as their implications and opportunities for future work.

\begin{table}
\label{tab:legacy}
\begin{center}
\caption{Legacy of current study}
\begin{tabular}{lc}
\hline
\hline
Development & Reference(s) \\
\hline
\hline
GCM benchmark tests & \cite{heng11a,heng11b} \\
\hline
Improved two-stream & \cite{heng14,heng18} \\
radiative transfer theory & \cite{hk17} \\
\hline
Non-hydrostatic & \cite{mendonca16} \\
dynamical core$^\dagger$ & \cite{deitrick20} \\
\hline 
Equilibrium chemistry & \cite{ht16} \\
 & \cite{stock18} \\
\hline
Chemical relaxation$^\ddagger$ & \cite{tsai18} \\
(GCM disequilibrium chemistry) & \cite{mendonca18b} \\
\hline
Opacities & \cite{gh15} \\
 & \cite{grimm21} \\
\hline
Temperature-pressure profiles & \cite{heng12,heng14} \\
Two-stream radiative transfer & \cite{malik17,malik19} \\
Cloud properties & \cite{kh18} \\
\hline
\hline
\end{tabular}\\
\end{center}
{\scriptsize $\dagger$: In the current study, the terms ``dynamical core" and ``GCM" are used interchangeably, but strictly speaking the former refers only to the core part of the GCM that solves the coupled equations of fluid dynamics. \\
$\ddagger$: Results using chemical relaxation are not explicitly shown in the current study, but this capability is already built into \texttt{THOR}.}
\end{table}

\section{Methodology}
\label{sect:methods}

\subsection{Previous developments}

For the current paper to be self-contained, concise reviews of previous codes and techniques are provided, which also give context to the new developments presented.  The computer codes are publicly available at \texttt{https://www.github.com/exoclime}.

\subsubsection{\texttt{THOR} general circulation model}

\begin{figure}
\begin{center}
\includegraphics[width=0.7\columnwidth]{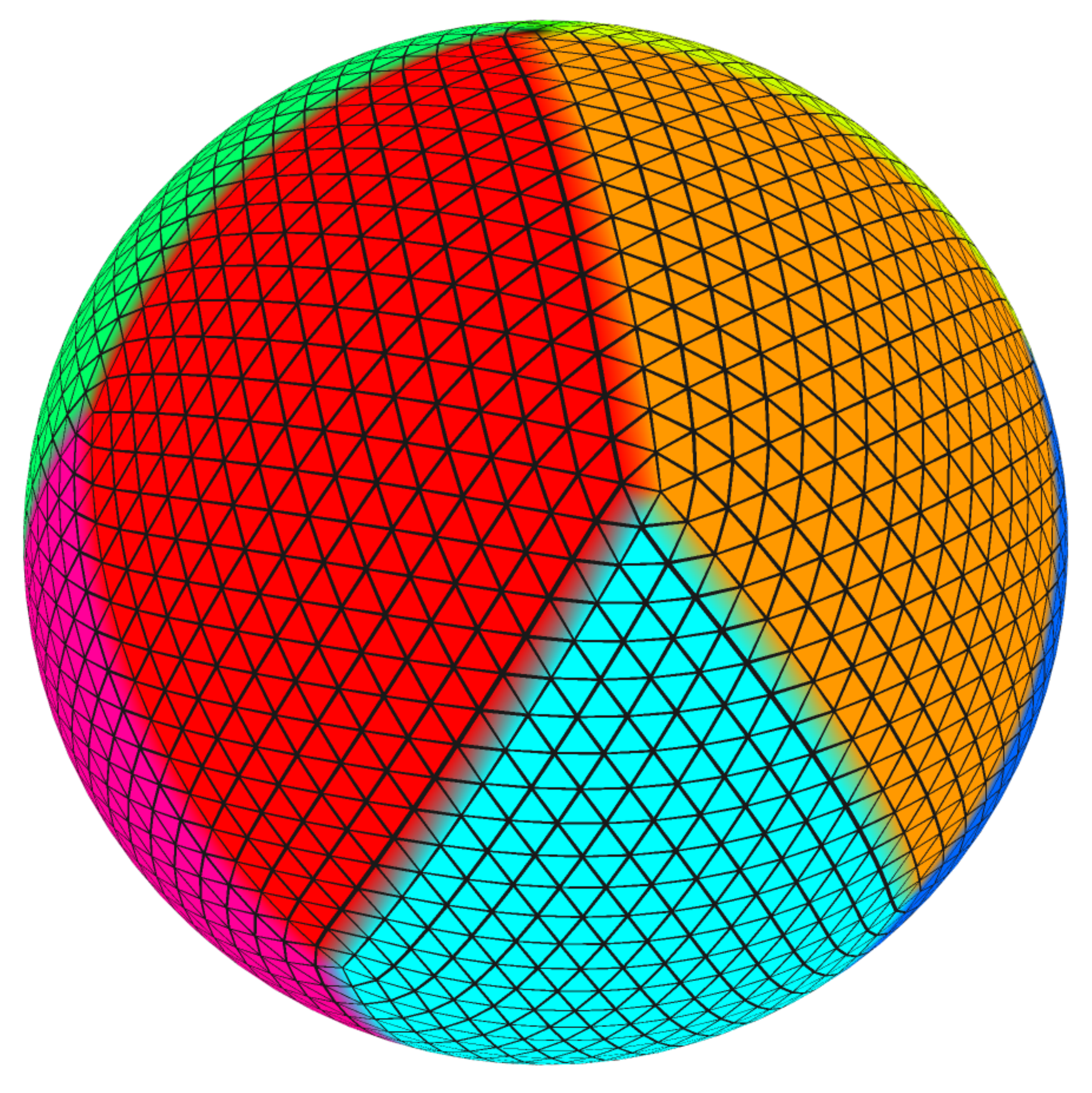}
\end{center}
\vspace{-0.1in}
\caption{Icosahedral grid used in \texttt{THOR}.  Each rhomboid represents regions on the grid that map to a square array of points, linearly stored in memory. Equations for points on a rhomboid are evaluated as a block in parallel on one core per point on the GPU.  This partitioning into memory chunks allows the GPU to parallelize operations on the grid more efficiently by requiring fewer memory access operations.}
\vspace{-0.1in}
\label{fig:horizontal_grid}
\end{figure}

\begin{figure*}
\begin{center}
\includegraphics[width=\textwidth]{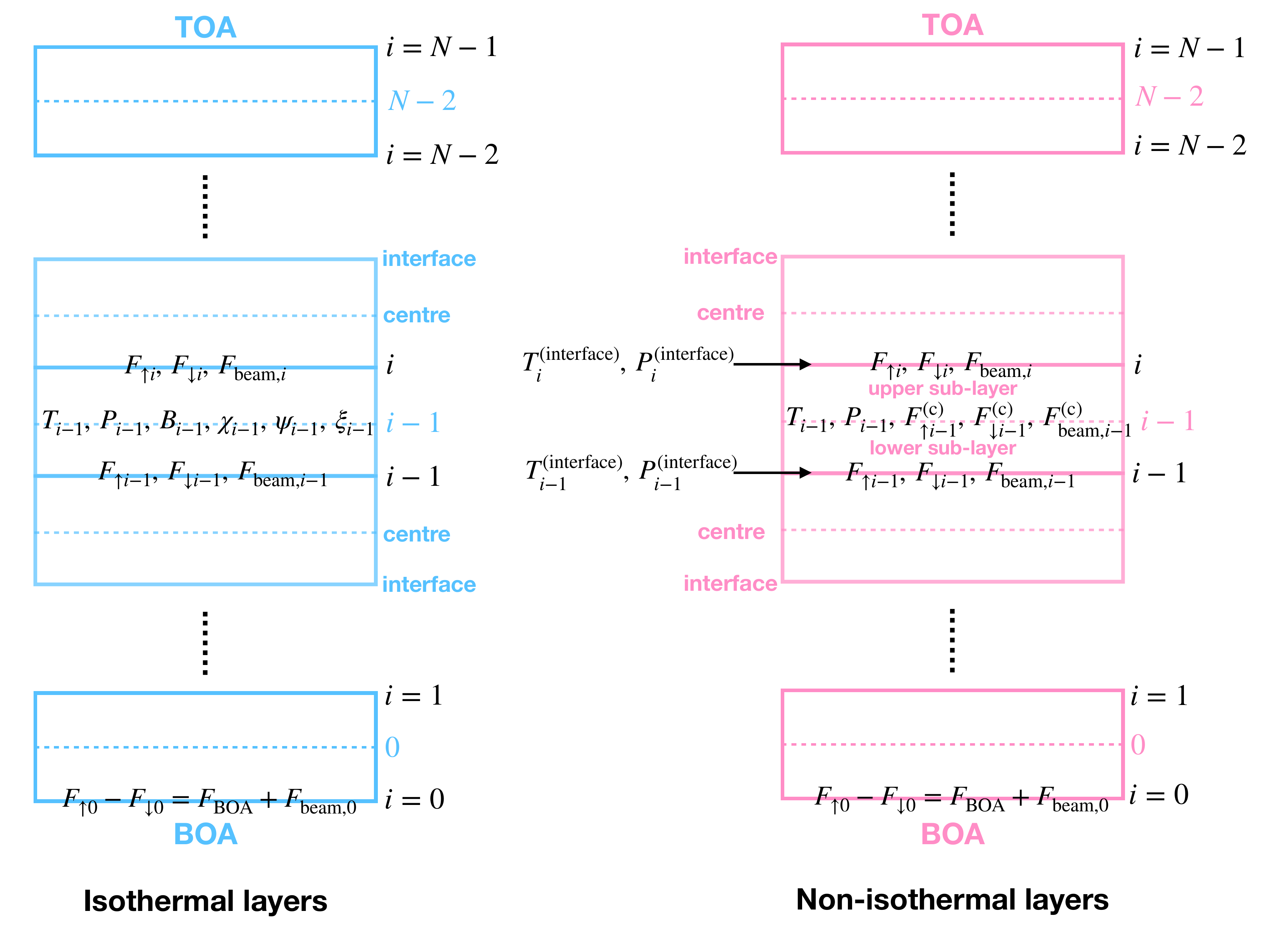}
\end{center}
\vspace{-0.1in}
\caption{Schematics of the staggered radial/vertical grid used in the \texttt{THOR+HELIOS} general circulation model.  The schematic on the left is for isothermal layers, where the temperature, pressure, Planck function, etc, are defined only at the center of each layer.  Both the stellar beam and diffuse fluxes exist only at the interfaces. The schematic on the right is for non-isothermal layers (non-constant Planck function), which are each divided into upper and lower sub-layers. Temperatures, pressures, fluxes, etc, exist at both the center of each layer and its interfaces. Quantities within each sub-layer take on values that are the average of the center and interface.  TOA and BOA are acronyms for top and bottom of atmosphere, respectively.}
\vspace{-0.1in}
\label{fig:grid}
\end{figure*}

The \texttt{THOR} GCM was the first to be developed from scratch for the study of exoplanetary atmospheres \citep{mendonca16}, rather than being adapted from GCMs used for Earth or solar system planets.  Unlike most other GCMs used for exoplanets, the dynamical core of \texttt{THOR} solves the full set of non-hydrostatic Euler equations, rather than the reduced set of primitive equations of meteorology (for a review, see Appendix \ref{append:review}).  The horizontally explicit vertically implicit (HEVI) algorithm used within \texttt{THOR} and implemented on an icosahedral grid with ``spring dynamics" (which allows the pentagons and hexagons of the grid to be projected onto spherical surfaces) is directly taken from the Earth science literature \citep{satoh02,satoh03,ts04,satoh08} and first implemented by \cite{mendonca16}.  Following \cite{deitrick20}, we use a horizontal spatial resolution of the icosahedral grid of $g_{\rm level}=4$, which corresponds to an angular resolution of about $4^\circ$ on the sphere.  When interpolated onto a latitude-longitude grid, this corresponds to 45 latitude and 90 longitude points.  In the vertical/radial direction, the grid has 40 spatial points.  It is worth noting that \texttt{THOR} uses MKS physical units to respect the convention of Earth GCMs. The vertical velocity is held at zero at the top and bottom boundaries of the model to conserve mass \citep{staniforth2003}. Near the upper boundary, there is a ``sponge-layer'', wherein the winds are damped toward their zonal means to mimic wave-breaking and prevent spurious reflections \citep{jablonowski11,mendonca18b}. 

Since \texttt{THOR} is a non-hydrostatic model, acoustic waves are permitted \citep{ts04,deitrick20}. However, these waves are very low in energy and have very little impact on circulation \citep{mendonca16}. Without using extremely small time steps, acoustic waves can be a source of noise, necessitating the use of divergence damping \citep{ts04,mendonca16}. 

Since hot Jupiter atmospheres can have supersonic winds, it has been suggested that shocks will form along the eastern terminator, where night-side equatorial winds collide with warmer, slower material \citep{goodman09,li10,dd10,heng12c}. However, 2-D shock-capturing simulations in \cite{fromang16} suggested that flows on hot Jupiters are smoothly decelerated through the sonic point. The \texttt{THOR} algorithm is not designed to capture shocks, thus we are not prepared to weigh-in on the matter of shocks. The computationally efficient, shock-capturing 3-D GCM introduced in \cite{ge20} may be well suited to address this issue.

\subsubsection{\texttt{HELIOS} radiative transfer code}

The \texttt{HELIOS} radiative transfer code \citep{malik17} is based on implementing the workhorse two-stream method \citep{schuster,mw80,toon89,pierrehumbert,heng14}.  A later version of the code \citep{malik19} implemented the improved two-stream radiative transfer method \citep{hk17,heng18} and dry convective adjustment \citep{manabe65}.  Equilibrium chemistry is computed using the \texttt{FastChem} code \citep{stock18}.  When used on its own, \texttt{HELIOS} performs an iteration to solve for radiative-convective equilibrium given assumptions about the chemistry of the atmosphere and its initial temperature-pressure profile.  When used in tandem with the \texttt{THOR} GCM, the iteration for radiative-convective equilibrium is deactivated.  This is because \texttt{THOR} iterates for a more general equilibrium between the three-dimensional atmospheric dynamics
and radiative heating.  Given a temperature-pressure profile supplied by \texttt{THOR}, \texttt{HELIOS} computes the radiative fluxes associated with heating/cooling, which are subsequently used to update the temperature-pressure profile (Figure \ref{fig:schematic}).

The original \texttt{HELIOS} code was written using both the \texttt{Python} and \texttt{CUDA C++} programming languages.  In order to perform the coupling to \texttt{THOR} without suffering a computational bottleneck, another version of \texttt{HELIOS} was written in \texttt{C++}; it is internally named \texttt{Alfrodull} for book-keeping purposes.\footnote{\texttt{https://www.github.com/exoclime/Alfrodull}}  It is worth noting that the original \texttt{HELIOS} code uses CGS physical units, whereas \texttt{Alfrodull} uses MKS physical units to be consistent with \texttt{THOR}.

\cite{malik17} benchmarked the \texttt{HELIOS} code against \cite{millerricci10}, finding that results for GJ 1214b compared well between the two models, both in the temperature-pressure profile and the spectrum. The T-P profile was produced using $k$-tables and the spectrum was produced using high-resolution opacity sampling. Further validation of \texttt{HELIOS} against other codes utilizing correlated$-k$, \texttt{Exo-REM} \citep{baudino2015}, \texttt{petitCODE} \citep{molliere2015}, and \texttt{ATMO} \citep{amundsen2014}, was provided in \cite{malik19}. 

\subsubsection{\texttt{HELIOS-K}: atmospheric opacity calculator}
\label{subsect:heliosk}

Atmospheric opacities (cross sections per unit mass) are calculated using the \texttt{HELIOS-K} calculator \citep{gh15,grimm21}.  Drawing from the \texttt{HITEMP} and \texttt{ExoMol} spectroscopic databases, we include contributions from several major carbon, oxygen, nitrogen and sulphur carriers: H$_2$O \citep{poly18}, CO \citep{li15}, CO$_2$ \citep{rothman10}, CH$_4$ \citep{y13,yt14}, NH$_3$ \citep{y11}, HCN \citep{harris06,barber14}, C$_2$H$_2$ \citep{chubb20}, PH$_3$ \citep{ss15} and H$_2$S \citep{azzam16}.  We also include collision-induced absorption due to H$_2$-H$_2$ \citep{abel11} and H$_2$-He \citep{abel12} pairs.  Truncated Voigt profiles with a line-wing cutoff of 100 cm$^{-1}$ are assumed.  Pressure broadening is included using standard broadening parameters provided by \texttt{HITEMP} and \texttt{ExoMol}.  All of the opacities used are publicly available at \texttt{https://dace.unige.ch/}.

For integration of the GCM, we utilize $k$-distributions with 30 bins, equally spaced in wavenumber from 0.3 $\mu$m to 200 $\mu$m.  We construct $k$-tables using the opacities sampled at the native resolution of \texttt{HELIOS-K}, which has a spacing in wavenumber of $\delta \nu = 0.01$ cm$^{-1}$. For this study, the k-table bins are equal size in wavenumber, and we use 30 bins (but see Section \ref{sect:benchmark}). The high-resolution opacities are combined (i.e., pre-mixed), across  temperature ($50 \le T \le 3000$ K; 60 values) and pressure ($10^{-6} \le P \le 10^3$ bar; 28 values equally spaced in $\log{P}$), assuming chemical equilibrium and a metallicity of [Fe/H]$=-0.13$ \citep[the metallicity of the host star;][]{sousa18}. Equilibrium chemistry calculations are performed using the \texttt{FastChem} code \citep{stock18}. Finally, within each bin, the resulting opacities are sorted and interpolated onto 20 Gaussian points \citep[see Equations 33 and 34 of][]{malik17}.  
Post-processing uses opacity sampling with a spectral resolution of 500, which corresponds to 3255 wavelength points. Spectral resolution is defined here to be
\begin{equation}
    R = \frac{\lambda}{\Delta \lambda},
\end{equation}
which results in a logarithmic spacing of samples.

\subsubsection{\texttt{LX-MIE} Mie scattering code: cloud/haze properties}
\label{subsect:lxmie}

To include the effects of clouds or hazes, one needs to compute the absorption and scattering cross sections of their constituent particles, as well as the scattering asymmetry factors (which describe how anisotropic or isotropic the scattered radiation is). To this end, we use the \texttt{LX-MIE} Mie scattering code \citep{kh18}.  We are agnostic about the formation mechanism of these constituent particles and term them ``aerosols" or ``condensates".  For the current study, these terms are used synonymously, because we do not attempt to model cloud or haze formation and only include their absorption and scattering effects on the radiative transfer.

\subsubsection{\texttt{PHOENIX} stellar template}

\begin{figure}
\begin{center}
\vspace{-0.1in}
\includegraphics[width=\columnwidth]{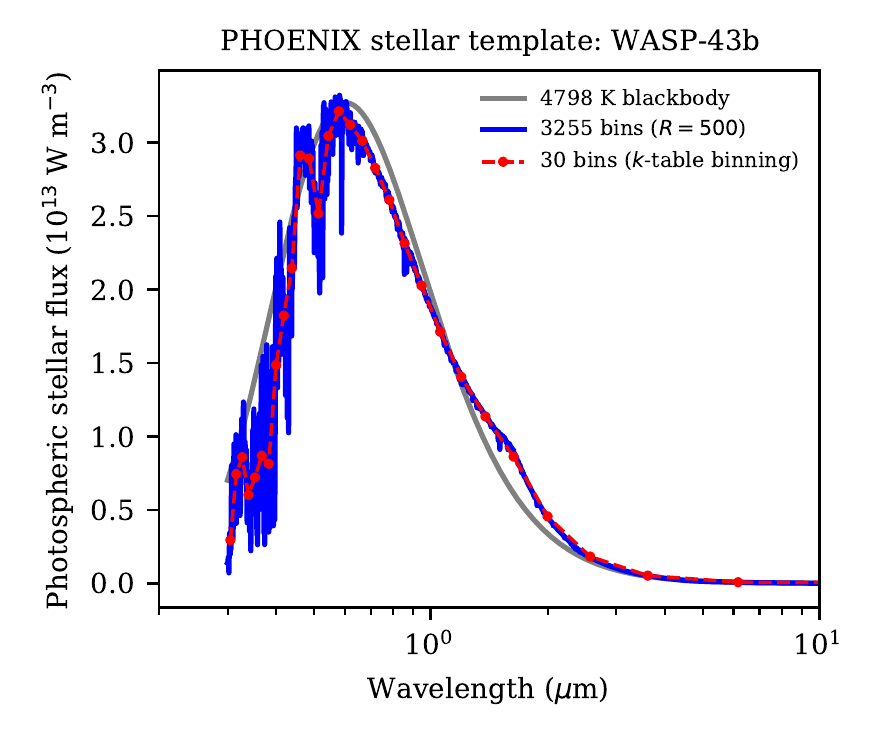}
\end{center}
\vspace{-0.1in}
\caption{\texttt{PHOENIX} stellar template of WASP-43 interpolated at two different spectral resolutions.  A 4798 K blackbody is shown for comparison. The blue curve is the high-resolution ($R=500$) spectrum used for post-processing. The red curve shows the spectrum used for integration; this spectrum is constructed from the original \texttt{PHOENIX} template by averaging the flux over each $k$-table bin (used for the opacities; see Section \ref{subsect:heliosk}). The red points indicate the centers of the $k$-table bins.}
\vspace{-0.1in}
\label{fig:stellar}
\end{figure}

In the current study, we focus on the hot Jupiter WASP-43b \citep{hellier11,gillon12}.  To model the stellar radiation incident upon this gas-giant exoplanet, we use stellar templates from the \texttt{PHOENIX} library \citep{husser2013}.  To interpolate for the stellar template of WASP-43, we use the following values of the stellar properties \citep{sousa18}: $T_\star = 4798$ K, $\log{g_\star}=4.55$ (cgs units), [Fe/H]$=-0.13$.  Given the lack of information, the alpha element enhancement is assumed to be solar.  For integration, we average the \texttt{PHOENIX} template over the opacity $k$-table bins, while for post-processing, we use opacity sampling at resolution $R=500$.  Figure \ref{fig:stellar} shows the final stellar template used.

\subsection{New developments: \texttt{THOR+HELIOS}}

\begin{figure*}
\begin{center}
\includegraphics[width=\textwidth]{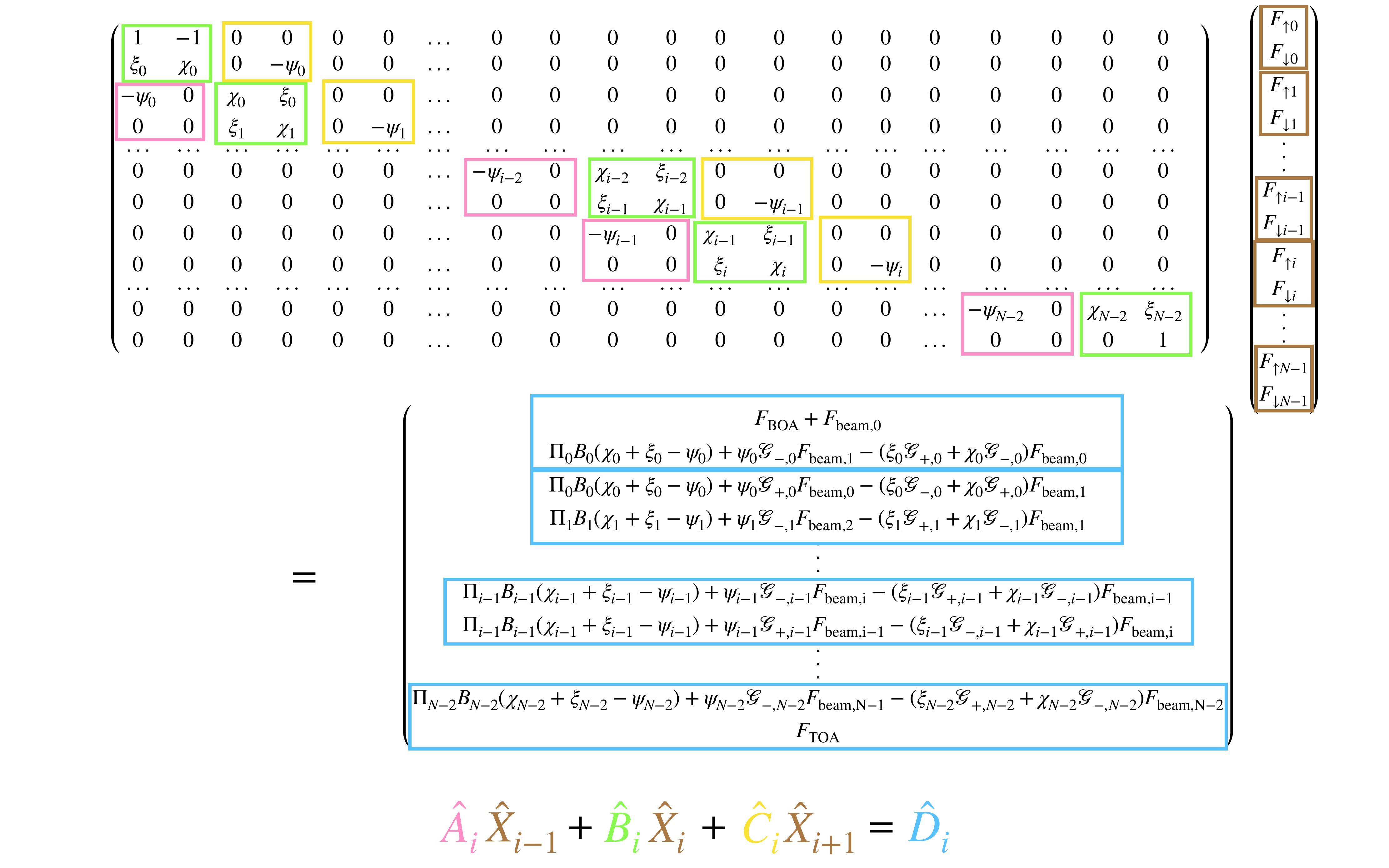}
\end{center}
\caption{Matrix form of the improved two-stream solutions of radiative transfer in the \texttt{THOR+HELIOS} general circulation model. The $2\times 1$ matrix $\hat{X}$, which contains the outgoing and incoming fluxes, may be obtained using the Thomas's algorithm (see text for details), where the tridiagonal matrix operating on $\hat{X}$ is composed of elements $\hat{A}$, $\hat{B}$ and $\hat{C}$ that are $2\times2$ matrices.  The $F_{\rm TOA}$ term refers to the diffuse infrared emission (and \textit{not} the stellar beam) at the top of the atmosphere, which is normally set to zero; it is available in the \texttt{THOR+HELIOS} code as an option for testing.  For exoplanets without surfaces, we set $F_{\rm BOA}=0$ as interior heating is included elsewhere in the algorithm.}
\label{fig:thomas_isothermal}
\end{figure*}

\subsubsection{Model grid, radiative transfer equations and boundary conditions}

\texttt{THOR+HELIOS} uses a staggered radial/vertical grid that is inherited from \texttt{HELIOS}.  It has two variants: isothermal and non-isothermal layers (Figure \ref{fig:grid}), which correspond to whether only a constant Planck function is assumed or its gradient is additionally computed, respectively \citep{heng14,malik17}.  The temperature and pressure at the center of each isothermal layer are provided by \texttt{THOR}, which are then used to compute the opacities, single-scattering albedos, mean molecular mass, Planck function, etc.  The stellar beam and diffuse fluxes exist only at the interfaces of each layer and are computed by \texttt{HELIOS}.  For non-isothermal layers, each layer is divided into upper and lower sub-layers \citep{mendonca15,malik17,malik19}.  
While temperatures and pressures at the center of the layer are already defined by the dynamical core, the values at the interfaces are determined by linear interpolation. Fluxes are defined at the interfaces and at the layer centers.
The values of the opacities, single-scattering albedos, mean molecular mass, Planck function, etc, are computed at both layer centers and interfaces; their values in the upper and lower sub-layers are taken to be the arithmetic mean of their central and interface values.

Appendix \ref{append:equations} reviews and re-casts the improved two-stream flux solutions of \cite{heng18} mostly in the notation of \cite{malik19}, which is the form implemented in the \texttt{HELIOS} code.  Consider an isothermal layer with center index $i-1$ and interface indices $i-1$ (lower interface) and $i$ (upper interface), as shown in Figure \ref{fig:grid}.  The outgoing (upward) flux at the upper interface is
\begin{equation}
\begin{split}
\chi_{i-1} F_{\uparrow i} =& \psi_{i-1} F_{\uparrow i-1} - \xi_{i-1} F_{\downarrow i} \\
&+ \Pi_{i-1} B_{i-1} \left( \chi_{i-1} + \xi_{i-1} - \psi_{i-1}\right) \\
&+ \psi_{i-1} {\cal G}_{+,i-1} F_{{\rm beam},i-1} \\
&- \left( \xi_{i-1} {\cal G}_{-,i-1} + \chi_{i-1} {\cal G}_{+,i-1} \right) F_{{\rm beam},i}.
\end{split}
\end{equation}
The incoming (downward) flux at the lower interface is
\begin{equation}
\begin{split}
\chi_{i-1} F_{\downarrow i-1} =& \psi_{i-1} F_{\downarrow i} - \xi_{i-1} F_{\uparrow i-1} \\
&+ \Pi_{i-1} B_{i-1} \left( \chi_{i-1} + \xi_{i-1} - \psi_{i-1} \right) \\
&+ \psi_{i-1} {\cal G}_{-,i-1} F_{{\rm beam},i} \\
&- \left( \xi_{i-1} {\cal G}_{+,i-1} + \chi_{i-1} {\cal G}_{-,i-1} \right) F_{{\rm beam},i-1}.
\end{split}
\end{equation}

For non-isothermal layers, the Planck function varies over the layer and it has a non-zero gradient with respect to the optical depth.  There are now four expressions for the fluxes (Figure \ref{fig:grid}).  Quantities at the center of each layer are superscripted by ``$(c)$"; in the upper and lower sub-layers, they are superscripted by ``(upper)" and ``(lower)", respectively.  The outgoing (upward) flux at the upper interface is
\begin{equation}
\begin{split}
\chi^{(\rm upper)}_{i-1} F_{\uparrow i} &= \psi^{(\rm upper)}_{i-1} F^{(c)}_{\uparrow i-1} - \xi^{(\rm upper)}_{i-1} F_{\downarrow i} \\
&+ \Pi^{(\rm upper)}_{i-1} B_i \left( \chi^{(\rm upper)}_{i-1} + \xi^{(\rm upper)}_{i-1} \right) \\
&- \Pi^{(\rm upper)}_{i-1} \psi^{(\rm upper)}_{i-1} B^{(c)}_{i-1} \\ 
&+ \Pi^{(\rm upper)}_{i-1} \frac{B^{(c)}_{i-1} - B_i}{\tau^{(c)}_{i-1} - \tau_i} ~\Theta^{(\rm upper)}_{i-1} \\
&+ \psi^{(\rm upper)}_{i-1} {\cal G}^{(\rm upper)}_{+,i-1} F^{(c)}_{{\rm beam},i-1} \\
&- \left( \xi^{(\rm upper)}_{i-1} {\cal G}^{(\rm upper)}_{-,i-1} + \chi^{(\rm upper)}_{i-1} {\cal G}^{(\rm upper)}_{+,i-1} \right) F_{{\rm beam},i}.
\end{split}
\label{eq:upflux_interface}
\end{equation}
where we have defined
\begin{equation}
\begin{split}
\Theta^{(\rm upper)}_{i-1} \equiv& \frac{1}{2E^{(\rm upper)}_{i-1} \left( 1 - \omega^{(\rm upper)}_{0,i-1} g^{(\rm upper)}_{0,i-1} \right)} \\
&\times \left( \chi^{(\rm upper)}_{i-1} - \psi^{(\rm upper)}_{i-1} - \xi^{(\rm upper)}_{i-1} \right).
\end{split}
\end{equation}
If $(\tau^{(c)}_{i-1} - \tau_i) < e_\tau$, then the gradient of the Planck function (fourth line of equation [\ref{eq:upflux_interface}]) is set to zero and the Planck functions (second and third lines of equation [\ref{eq:upflux_interface}]) are replaced by $(B^{(c)}_{i-1} + B_i)/2$.  The tolerance $e_\tau=10^{-4}$ allows a non-isothermal layer to become an isothermal one when the optical depth of the layer is too small and ensures numerical stability.  The incoming (downward) flux at the center of the layer also uses quantities from the upper sub-layer,
\begin{equation}
\begin{split}
\chi^{(\rm upper)}_{i-1} F^{(c)}_{\downarrow i-1} &= \psi^{(\rm upper)}_{i-1} F_{\downarrow i} - \xi^{(\rm upper)}_{i-1} F^{(c)}_{\uparrow i-1} \\
&+ \Pi^{(\rm upper)}_{i-1} B^{(c)}_{i-1} \left( \chi^{(\rm upper)}_{i-1} + \xi^{(\rm upper)}_{i-1} \right) \\
&- \Pi^{(\rm upper)}_{i-1} \psi^{(\rm upper)}_{i-1} B_i \\ 
&- \Pi^{(\rm upper)}_{i-1} \frac{B^{(c)}_{i-1} - B_i}{\tau^{(c)}_{i-1} - \tau_i} ~\Theta^{(\rm upper)}_{i-1} \\
&+ \psi^{(\rm upper)}_{i-1} {\cal G}^{(\rm upper)}_{-,i-1} F_{{\rm beam},i} \\
&- \left( \xi^{(\rm upper)}_{i-1} {\cal G}^{(\rm upper)}_{+,i-1} + \chi^{(\rm upper)}_{i-1} {\cal G}^{(\rm upper)}_{-,i-1} \right) F^{(c)}_{{\rm beam},i-1}.
\end{split}
\label{eq:downflux_center}
\end{equation}
The other two fluxes use quantities from the lower sub-layer.  The outgoing (upward) flux at the center of the layer is
\begin{equation}
\begin{split}
\chi^{(\rm lower)}_{i-1} F^{(c)}_{\uparrow i-1} &= \psi^{(\rm lower)}_{i-1} F_{\uparrow i-1} - \xi^{(\rm lower)}_{i-1} F^{(c)}_{\downarrow i-1} \\
&+ \Pi^{(\rm lower)}_{i-1} B^{(c)}_{i-1} \left( \chi^{(\rm lower)}_{i-1} + \xi^{(\rm lower)}_{i-1} \right) \\
&- \Pi^{(\rm lower)}_{i-1} \psi^{(\rm lower)}_{i-1} B_{i-1} \\ 
&+ \Pi^{(\rm lower)}_{i-1} \frac{B_{i-1} - B^{(c)}_{i-1}}{\tau_{i-1} - \tau^{(c)}_{i-1}} ~\Theta^{(\rm lower)}_{i-1} \\
&+ \psi^{(\rm lower)}_{i-1} {\cal G}^{(\rm lower)}_{+,i-1} F_{{\rm beam},i-1} \\
&- \left( \xi^{(\rm lower)}_{i-1} {\cal G}^{(\rm lower)}_{-,i-1} + \chi^{(\rm lower)}_{i-1} {\cal G}^{(\rm lower)}_{+,i-1} \right) F^{(c)}_{{\rm beam},i-1},
\end{split}
\label{eq:upflux_center}
\end{equation}
where we have defined 
\begin{equation}
\begin{split}
\Theta^{(\rm lower)}_{i-1} \equiv& \frac{1}{2E^{(\rm lower)}_{i-1} \left( 1 - \omega^{(\rm lower)}_{0,i-1} g^{(\rm lower)}_{0,i-1} \right)} \\
&\times \left( \chi^{(\rm lower)}_{i-1} - \psi^{(\rm lower)}_{i-1} - \xi^{(\rm lower)}_{i-1} \right).
\end{split}
\end{equation}
If $(\tau_{i-1} - \tau^{(c)}_{i-1}) < e_\tau$, then the gradient of the Planck function (fourth line of equation [\ref{eq:upflux_center}]) is set to zero and the Planck functions (second and third lines of equation [\ref{eq:upflux_center}]) are replaced by $(B_{i-1} + B^{(c)}_{i-1})/2$.  Finally, the incoming (downward) flux at the lower interface is
\begin{equation}
\begin{split}
\chi^{(\rm lower)}_{i-1} F_{\downarrow i-1} &= \psi^{(\rm lower)}_{i-1} F^{(c)}_{\downarrow i-1} - \xi^{(\rm lower)}_{i-1} F_{\uparrow i-1} \\
&+ \Pi^{(\rm lower)}_{i-1} B_{i-1} \left( \chi^{(\rm lower)}_{i-1} + \xi^{(\rm lower)}_{i-1} \right) \\
&- \Pi^{(\rm lower)}_{i-1} \psi^{(\rm lower)}_{i-1} B^{(c)}_{i-1} \\ 
&- \Pi^{(\rm lower)}_{i-1} \frac{B_{i-1} - B^{(c)}_{i-1}}{\tau_{i-1} - \tau^{(c)}_{i-1}} ~\Theta^{(\rm lower)}_{i-1} \\
&+ \psi^{(\rm lower)}_{i-1} {\cal G}^{(\rm lower)}_{-,i-1} F^{(c)}_{{\rm beam},i-1} \\
&- \left( \xi^{(\rm lower)}_{i-1} {\cal G}^{(\rm lower)}_{+,i-1} + \chi^{(\rm lower)}_{i-1} {\cal G}^{(\rm lower)}_{-,i-1} \right) F_{{\rm beam},i-1}.
\end{split}
\label{eq:downflux_interface}
\end{equation}

\begin{figure*}
\begin{center}
\vspace{0.2in}
\includegraphics[width=\textwidth]{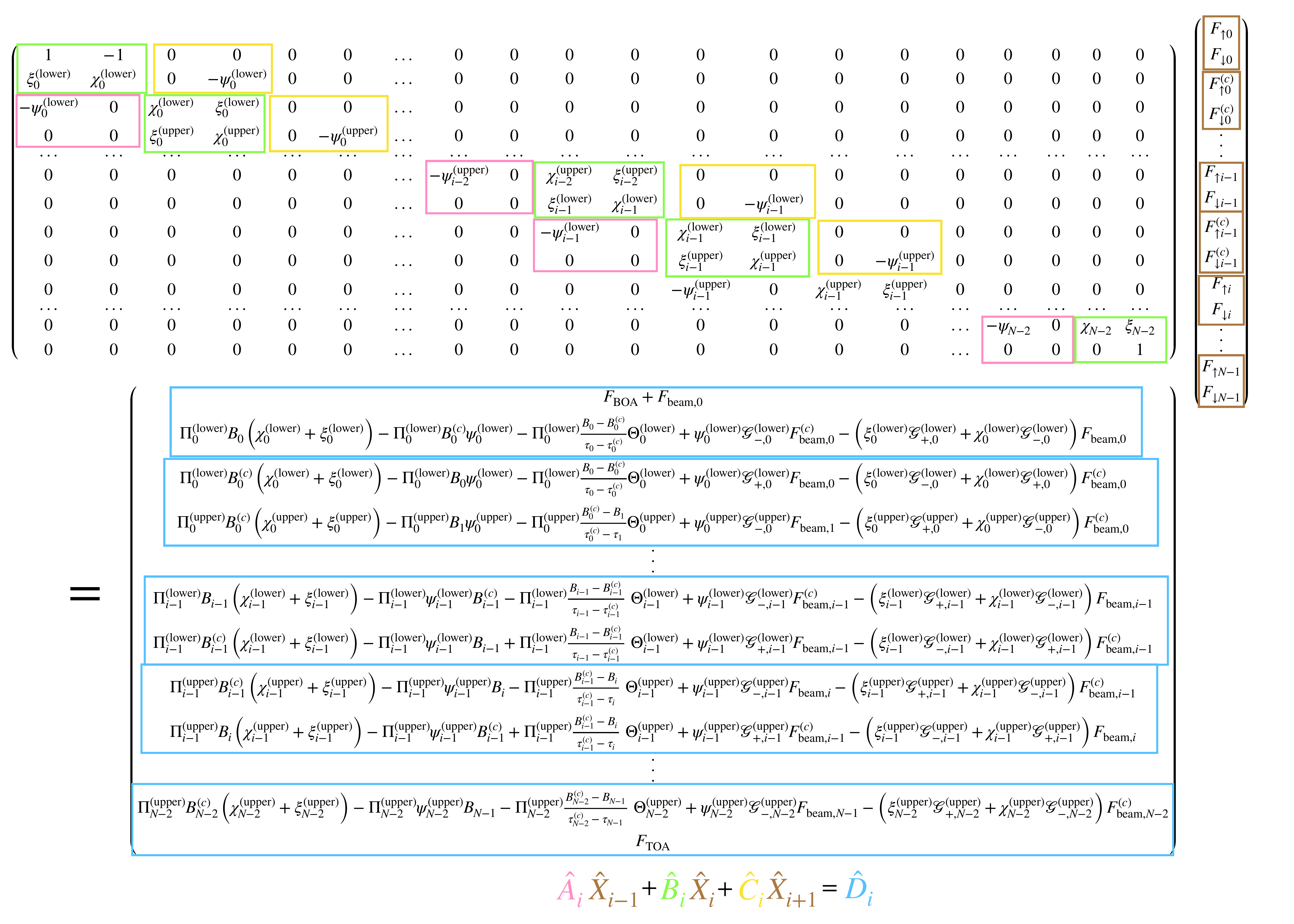}
\end{center}
\caption{Same as Figure \ref{fig:thomas_isothermal}, but for non-isothermal layers, where both the Planck function and its gradient are computed within each atmospheric layer.}
\vspace{0.1in}
\label{fig:thomas_non_isothermal}
\end{figure*}

\begin{figure}
\begin{center}
\vspace{-0.1in}
\includegraphics[width=\columnwidth]{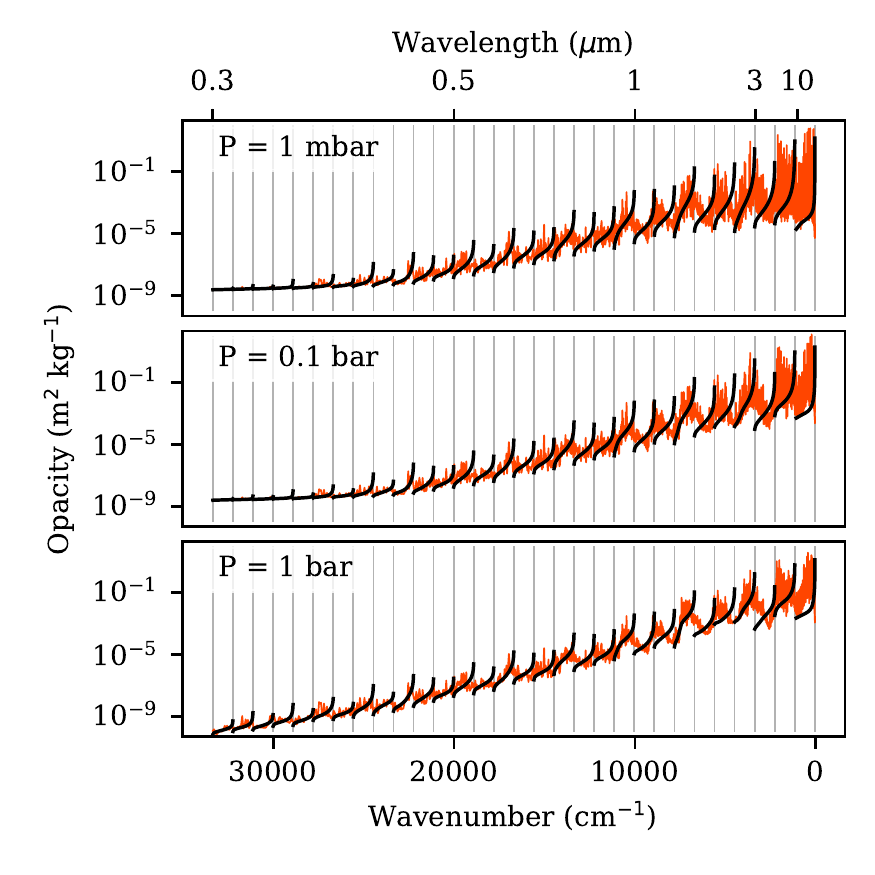}
\end{center}
\vspace{-0.2in}
\caption{Examples of combined opacity function, where individual molecular opacities of H$_2$O, CO, CO$_2$, CH$_4$, NH$_3$, HCN, C$_2$H$_2$, PH$_3$ and H$_2$S are weighted by their volume mixing ratios computed assuming chemical equilibrium. Each panel represents a different pressure at a temperature of $1500$ K.  The Rayleigh scattering cross sections enter into two-stream radiative transfer via the single-scattering albedo and the total optical depth.  The orange curves are the high-resolution ($R = 500$; 3255 wavelength points) opacity data that is used only for post-processing. Black curves represent the k-distribution tables (30 bins with 20 Gaussian points each) that are used during integration. Vertical gray lines delineate the k-table bins.}
\vspace{-0.1in}
\label{fig:opacity}
\end{figure}

\begin{figure}
\begin{center}
\vspace{-0.1in}
\includegraphics[width=\columnwidth]{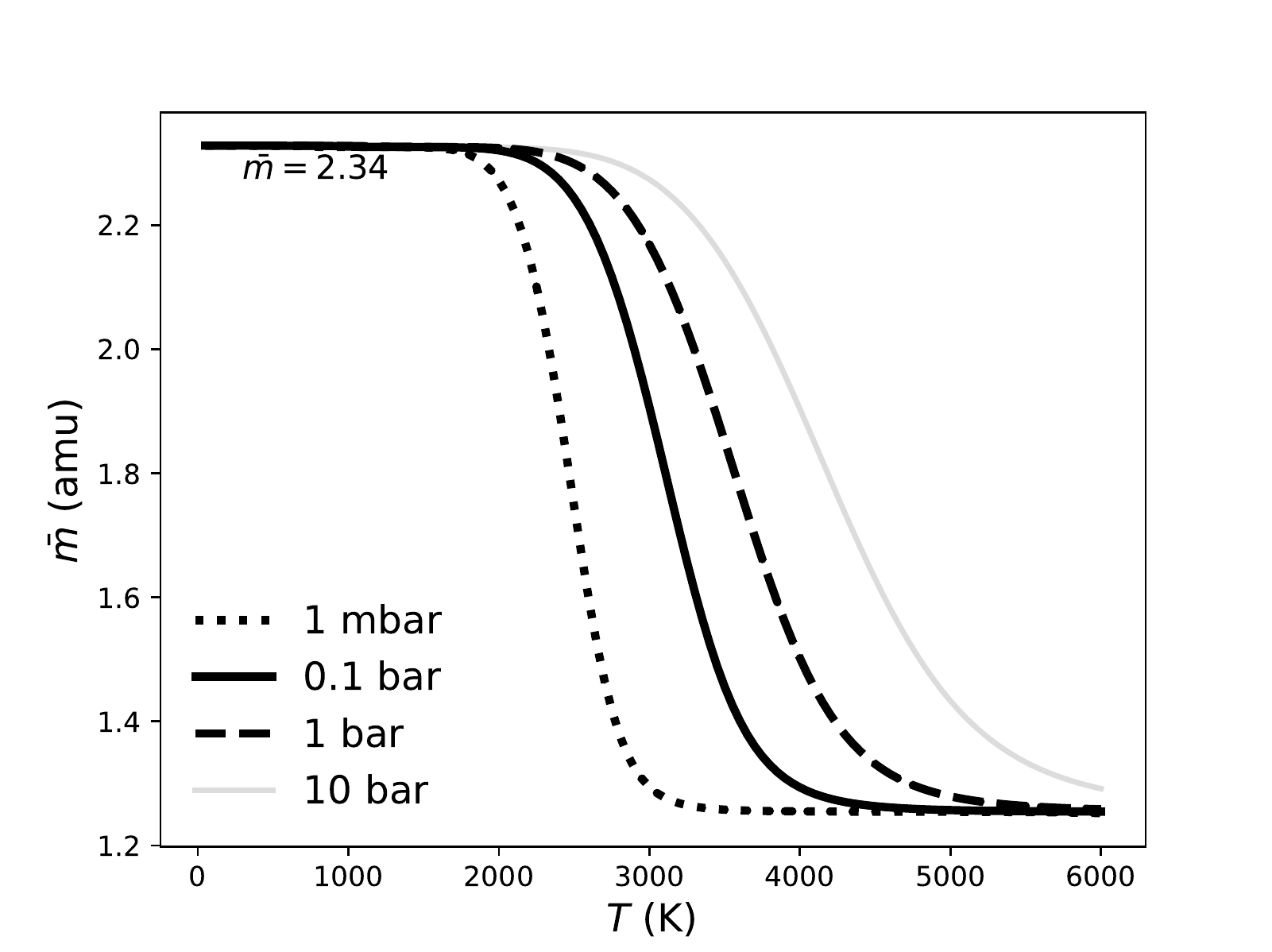}
\end{center}
\vspace{-0.2in}
\caption{Mean molecular mass as a function of temperature and pressure corresponding to the equilibrium chemistry model used in the current study.  At temperatures below 2000 K, the assumption of a constant mean molecular mass of $\bar{m}=2.34$ amu, corresponding to a gas dominated by molecular hydrogen, is reasonable.}
\vspace{-0.1in}
\label{fig:mm}
\end{figure}

\begin{figure}
\begin{center}
\vspace{-0.1in}
\includegraphics[width=\columnwidth]{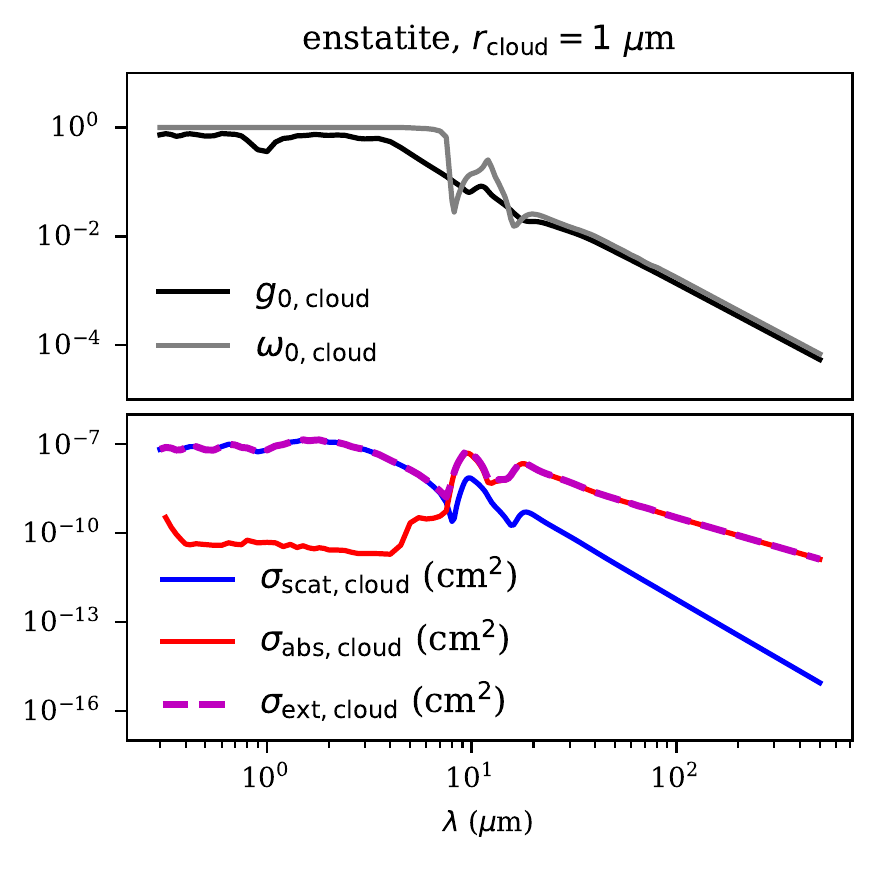}
\end{center}
\vspace{-0.2in}
\caption{Absorption cross section ($\sigma_{\text{abs}}$), scattering cross section ($\sigma_{\text{scat}}$), total extinction cross section ($\sigma_{\text{ext}}$), single scattering albedo ($\omega_0$), and scattering asymmetry factor ($g_0$) of spherical enstatite particles. These follow a monodisperse size distribution with a radius of $r_{\rm cloud}=1$ $\mu$m. For the current work, we assume these quantities are independent of temperature and pressure.}
\vspace{-0.1in}
\label{fig:enstatite}
\end{figure}

\begin{figure}
\begin{center}
\includegraphics[width=\columnwidth]{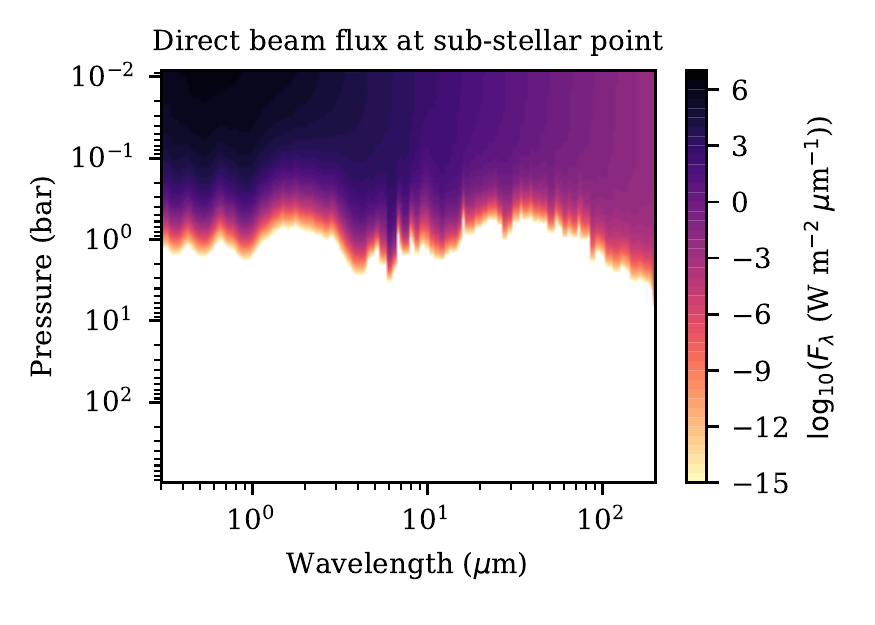}
\end{center}
\vspace{-0.2in}
\caption{Flux of the direct stellar beam as a function of both wavelength and pressure.  For illustration, we show calculations from the GCM of WASP-43b with non-isothermal layers and improved two-stream radiative transfer, but checked that the qualitative behaviour is similar for the other three GCMs (not shown).  Fluxes below $10^{-15}$ W m$^{-2}$ $\mu$m$^{-1}$ are assumed to be zero in this plot.}
\vspace{-0.1in}
\label{fig:beam}
\end{figure}

Heating of the atmosphere by the stellar beam is given by equation (\ref{eq:stellar_beam}), which is essentially Beer's law applied in the shortwave. Operationally, the flux associated with the stellar beam is added to the downward flux before the net flux is computed.  As the stellar beam is attenuated, it becomes diffuse emission. At the top boundary of the atmosphere, the direct beam enters at full strength; the upward stream of the diffuse beam escapes to space. At the bottom of the atmosphere (BOA), the boundary condition is
\begin{equation}
F_{\uparrow 0} - F_{\downarrow 0} = F_{\rm BOA} + F_{{\rm beam},0},
\end{equation}
where $F_{\rm BOA} = \pi B(T_{\rm int})$, $B$ is the Planck function and $T_{\rm int}$ is the interior temperature.  If the stellar beam is not attenuated at the BOA, then we have $F_{{\rm beam},0} \ne 0$ and it is artificially reflected upwards as part of the BOA boundary condition.  Examining the profile of $F_{{\rm beam},i}$ with radial distance is a sanity check to ensure that the BOA is located at a sufficiently deep pressure and/or if the adequate opacity sources have been included such that the model atmosphere is not implausibly transparent to starlight.

Inspection of equation (\ref{eq:g_factors}) reveals that there exists a critical value of $\mu_\star$ for which ${\cal G}_\pm$ diverges,
\begin{equation}
\mu_{\star,{\rm crit}} = \frac{1}{2\sqrt{E\left(E-\omega_0\right)\left(1-\omega_0 g_0 \right)}}.
\end{equation}
This issue has previously been noted and addressed by \cite{toon89}, who proposed that ``this problem can be eliminated by simply choosing a slightly different value of $\mu_\star$". In the 3-D simulations, this singularity is highly likely to appear and using this proposed solution can lead to unphysical forcing patterns. Instead, we perform a check on term associated with ${\cal G}_\pm$, $4 E \mu_{\star}^2 (E-\omega_0)(1-\omega_0 g_0) - 1$. It can be shown that as $\mu_{\star}$ approaches the critical value, this term reduces to $2 \mu_\star$. In the case that the full term is less than $10^{-5}$, we switch to this reduced form. It is not clear what the tolerance should be here; we have chosen a value that avoids the singularity effectively without causing unusual behavior.


\subsubsection{Multiple scattering using Thomas's algorithm}
\label{subsect:thomas}

The improved two-stream flux solutions, described in Appendix \ref{append:equations}, allow for radiation from an atmospheric layer to be scattered to its immediate neighbours.  In the absence of scattering, the arrays of outgoing and incoming fluxes may be computed independently of each other, because they depend only on the fluxes impinging upon the bottom and top of each layer, respectively.  When scattering is present, the outgoing or incoming flux of each layer now depends on both boundary conditions.  One may use an iterative approach to populate these flux arrays \citep{oreshenko16}.

When the calculation is repeated, radiation is scattered twice and may propagate across two layers.  When repeated ${\cal N}$ times, radiation is scattered ${\cal N}$ times in both directions---the multiple scattering of radiation in a model atmosphere.  The simplest implementation of multiple scattering is simply to iterate the two-stream solutions, across the entire atmosphere, for a finite number of times.  This is the approach adopted in the stand-alone \texttt{HELIOS} code; \cite{malik19} performed ${\cal N}=200$ iterations for multiple scattering.  The simplified GCMs of \cite{oreshenko16} also used this approach, typically enforcing $\sim 10$ iterations.

Instead of iterating pairwise through the entire atmosphere, a better approach is to cast the entire set of two-stream flux solutions in matrix form (Figure \ref{fig:thomas_isothermal}) and solve the system by matrix inversion. This approach has been used in numerous radiative-transfer models since its introduction in \cite{toon89}. We include a complete description of the method here since it is a new addition to the \texttt{HELIOS} model. 

\begin{figure*}
\begin{center}
\vspace{-0.1in}
\includegraphics[width=\textwidth]{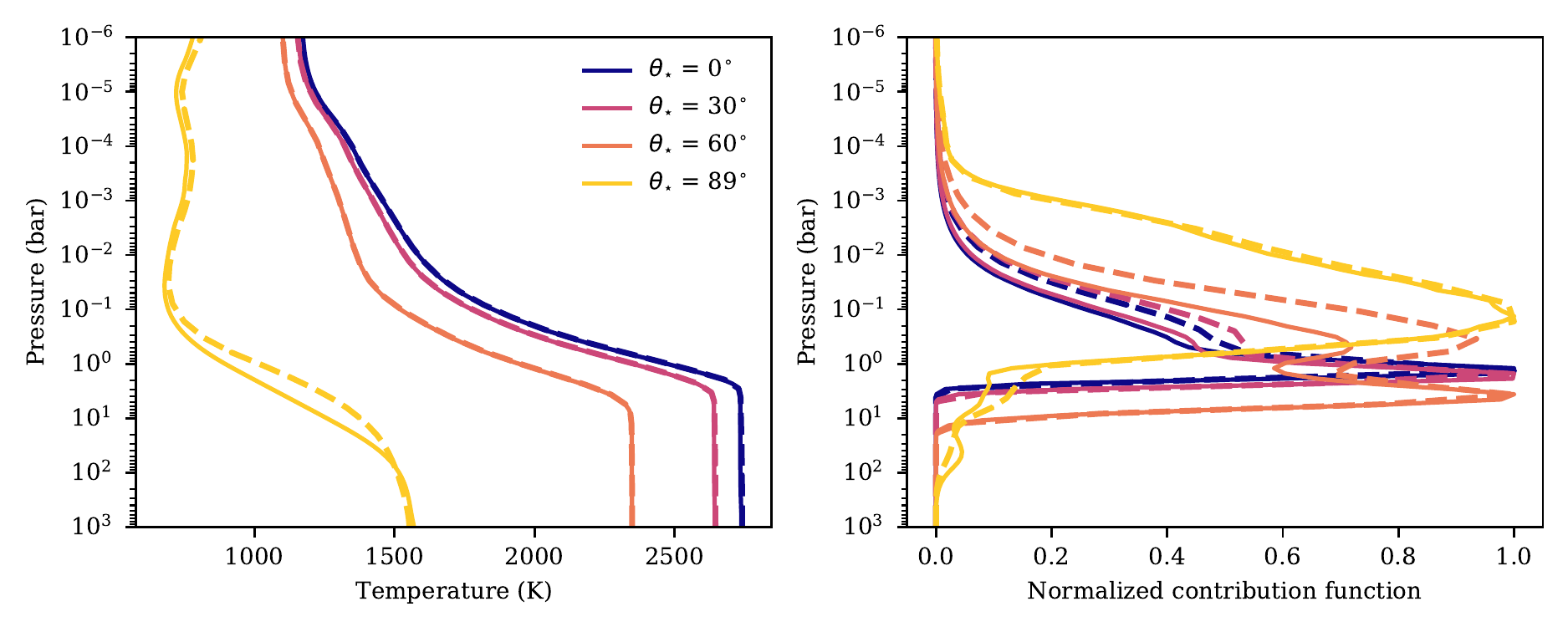}
\end{center}
\vspace{-0.2in}
\caption{\emph{Left:} Comparing \texttt{THOR+HELIOS} in 1-D mode against 1-D \texttt{HELIOS}, for 4 different zenith angles ($\theta_{star}$). The solid curves are from 1-D \texttt{HELIOS}, the dashed from \texttt{THOR+HELIOS}. \emph{Right:} the thermal contribution function (normalized to its peak value) from 1-D \texttt{HELIOS} (solid) and \texttt{THOR+HELIOS} in 1-D mode (dashed). }
\vspace{-0.1in}
\label{fig:helios_thor_comp}
\end{figure*}

\begin{figure*}
\begin{center}
\vspace{-0.1in}
\includegraphics[width=\textwidth]{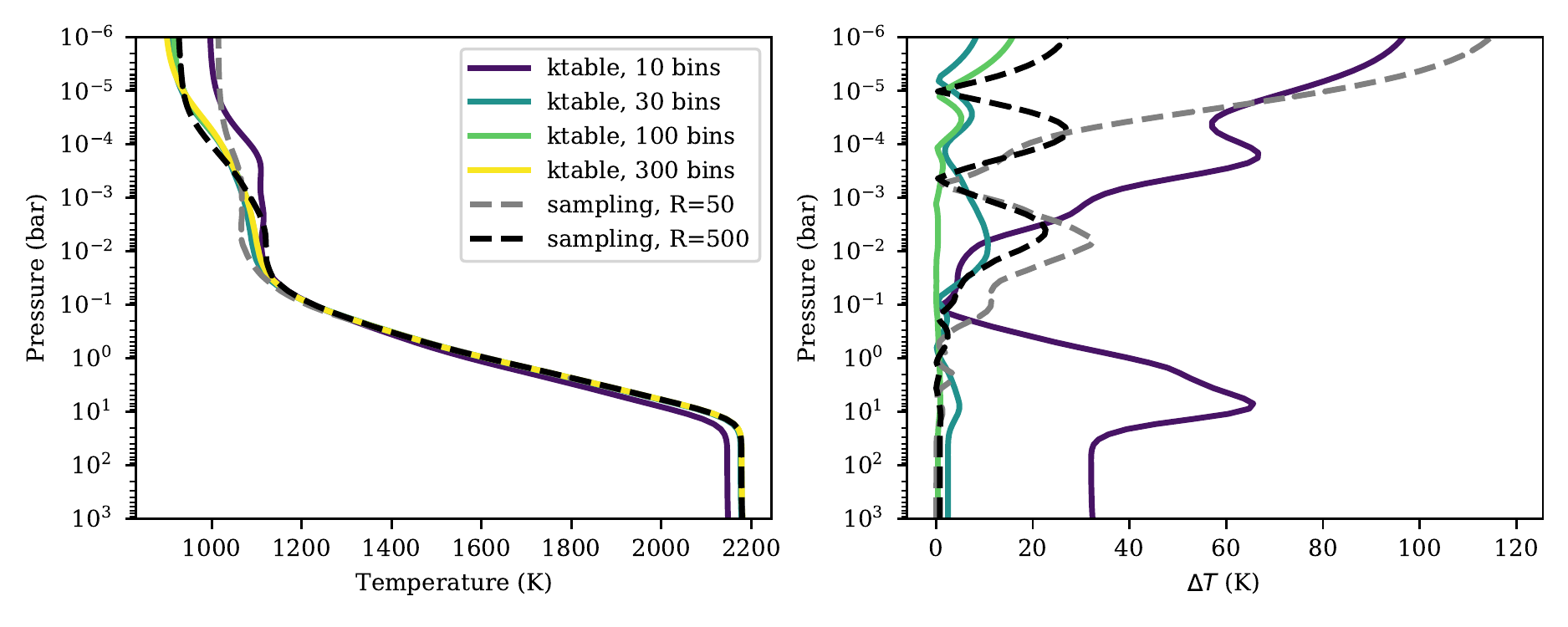}
\end{center}
\vspace{-0.2in}
\caption{ Comparing 1-D \texttt{HELIOS} simulations of WASP-43b at different spectral resolutions. \emph{Left:} Temperature-pressure profiles. \emph{Right:} Residuals in temperature for each simulation, compared against the 300 k-table bin simulation. }
\vspace{-0.1in}
\label{fig:helios_specres_test}
\end{figure*}

The matrices themselves contain $2\times2$ and $2\times1$ sub-matrices, which obey the following relation,
\begin{equation}
\hat{A}_i \hat{X}_{i-1} + \hat{B}_i \hat{X}_i + \hat{C}_i \hat{X}_{i+1} = \hat{D}_i,
\end{equation}
where the $2\times1$ sub-matrix $\hat{X}$ contains the outgoing and incoming fluxes.  The $2\times2$ sub-matrices $\hat{A}$, $\hat{B}$ and $\hat{C}$ contain the coefficients $\chi$, $\xi$ and $\psi$ (see equation [\ref{eq:coefficients}] for definitions).  The $2\times1$ sub-matrix $\hat{D}$ contains the blackbody and stellar beam terms.  Solving for $\hat{X}$ involves inverting a tridiagonal matrix where the elements are the sub-matrices $\hat{A}$, $\hat{B}$ and $\hat{C}$, which may be accomplished using Thomas's algorithm (e.g., Chapter 6.3 of \citealt{mihalas}), which first computes 
\begin{equation}
\hat{C}_i^\prime =
\begin{cases}
\hat{C}_0/\hat{B}_0, & i=0, \\
\hat{C}_i/\left( \hat{B}_i - \hat{A}_i \hat{C}_{i-1}^\prime \right), & \mbox{otherwise}, \\
\end{cases}
\end{equation}
\begin{equation}
\hat{D}_i^\prime =
\begin{cases}
\hat{D}_0/\hat{B}_0, & i=0, \\
\left( \hat{D}_i - \hat{A}_i \hat{D}_{i-1}^\prime \right)/\left( \hat{B}_i - \hat{A}_i \hat{C}_{i-1}^\prime \right), & \mbox{otherwise}, \\
\end{cases}
\end{equation}
followed by performing back-substitution,
\begin{equation}
\hat{X}_i =
\begin{cases}
\hat{D}_{N-1}^\prime, & i=N-1, \\
\hat{D}_i^\prime - \hat{C}_i^\prime \hat{X}_{i+1}, & \mbox{otherwise}. \\
\end{cases}
\end{equation}

For non-isothermal atmospheric layers, the elements of the matrices are different, but the procedure is conceptually identical (Figure \ref{fig:thomas_non_isothermal}).

During each iteration with \texttt{THOR} (Figure \ref{fig:schematic}), \texttt{HELIOS} uses this procedure to implement multiple scattering of radiation throughout the model atmosphere.

\subsubsection{Single-scattering albedo and scattering asymmetry factor}

\begin{table*}
\label{tab:parameters}
\begin{center}
\caption{List of input parameters}
\begin{tabular}{lcccc}
\hline
\hline
Name & Symbol & Value & Purpose & Reference \\
\hline
\hline
Acceleration due to gravity & $g$ & 47 m s$^{-2}$ & GCM & \cite{gillon12} \\
Radius$^\dagger$ & $R_p$ & $1.036 ~R_{\rm J} = 7.41 \times 10^7$ m& GCM & \cite{gillon12} \\
Reference pressure$^\dagger$ & $P_{\rm ref}$ & $1 \times 10^8$ Pa $=1000$ bar & GCM & --- \\
Altitude at top of simulation domain & --- & $1.7 \times 10^6$ m & GCM & --- \\
Angular rotational frequency & $\Omega$ & $8.94 \times 10^{-5}$ s$^{-1}$ & GCM & \cite{gillon12} \\
Specific gas constant & ${\cal R}$ & 3553 J kg$^{-1}$ K$^{-1}$ & GCM & $\clubsuit$ \\
Specific heat capacity & $c_P$ & 12436 J kg$^{-1}$ K$^{-1}$ & GCM & $\clubsuit$ \\
Initialisation temperature & $T_{\rm eq} = T_{\rm irr}/\sqrt{2}$ & 1725 K & GCM & \cite{gillon12} \\
Cloud-to-gas ratio (by number) & $f_{\rm cloud}$ & $10^{-17}$ & RT & --- \\
Semi-major axis & $a$ & 0.01525 AU$=2.28 \times 10^9$ m & RT & \cite{gillon12} \\
Stellar radius & $R_\star$ & $0.667 R_\odot = 4.64 \times 10^8$ m & RT & \cite{gillon12} \\
Stellar effective temperature & $T_\star$ & 4798 K & RT, stellar template & \cite{sousa18} \\
Stellar gravity & $\log{g_\star}$ & 4.55 (cgs units) & stellar template & \cite{sousa18} \\
Stellar metallcity & [Fe/H] & -0.13 & stellar template & \cite{sousa18} \\
Alpha element enhancement & [$\alpha$/M] & 0 (solar value) & stellar template & --- \\
Direct stellar beam Eddington coefficient & $\epsilon_2$ & 2/3 & RT & \cite{heng18} \\
\hline
\hline
\end{tabular}\\
\end{center}
$\dagger$: At bottom of simulation domain. \\
$\clubsuit$: Based on assuming 5 degrees of freedom (diatomic gas without vibrational modes activated) and $\bar{m}=2.34$. \\
Note: GCM refers to the dynamical core, RT stands for ``radiative transfer".
\end{table*}
Generally, radiation is scattered by both atoms/molecules and aerosols/condensates.  For the scattering cross section ($\sigma_{\rm scat, gas}$) associated with the gas, we consider Rayleigh scattering by CO$_2$, CO, H$_2$O, H, H$_2$ and He (Appendix \ref{append:rayleigh}).  For gas absorption, we use the \texttt{HELIOS-K} calculator to compute molecular opacities (Section \ref{subsect:heliosk}), which are then combined into a total absorption opacity (cross section per unit mass),
\begin{equation}
\kappa = \sum_i ~\kappa_i ~\frac{X_i m_i}{\bar{m}},
\end{equation}
where the sum is over all of the molecules considered, $\kappa_i$ is the opacity of the $i$-th molecule, $X_i$ is its volume mixing ratio, $m_i$ is its mass and $\bar{m}$ is the mean molecular mass of the gas.  Figure \ref{fig:opacity} shows examples of the combined opacity function.  Figure \ref{fig:mm} shows that $\bar{m}=2.34$ atomic mass units (amu) is a good approximation for most of the temperatures and pressures simulated by the GCM, which assumes $c_P \propto 1/\bar{m}$ to be constant.

For the absorption ($\sigma_{\rm abs, cloud}$) and scattering ($\sigma_{\rm scat, cloud}$) cross sections, as well as the scattering asymmetry factor ($g_{\rm 0, cloud}$), associated with aerosols/condensates, we use \texttt{LX-MIE} (Section \ref{subsect:lxmie}) to compute them for enstatite particles assuming a monodisperse size distribution with a  radius of 1 $\mu$m (Figure \ref{fig:enstatite}). 

The total single-scattering albedo, including gas and condensates, is constructed by weighing the terms by their respective number densities,
\begin{equation}
\omega_0 = \frac{\sigma_{\rm scat, gas} + f_{\rm cloud} \sigma_{\rm scat,cloud}}{\sigma_{\rm scat, gas} + \kappa \bar{m} + f_{\rm cloud} \left( \sigma_{\rm abs, cloud} + \sigma_{\rm scat, cloud} \right)}. \label{eqn:omega0}
\end{equation}
Similarly, the scattering asymmetry factor is
\begin{equation}
g_0 = \frac{f_{\rm cloud} \sigma_{\rm scat,cloud}}{\sigma_{\rm scat, gas} + f_{\rm cloud} \sigma_{\rm scat, cloud}} ~g_{\rm 0, cloud}. \label{eqn:g0}
\end{equation}
Since atoms/molecules have sizes that are much smaller than the optical/visible and infrared wavelengths considered, their scattering asymmetry factors are taken to be zero.  

The factor $f_{\rm cloud}$ is the ratio of number densities of the cloud to the gas.  In general, it is a function of temperature and pressure, and varies throughout the model atmosphere.  In the current study, we assume that $f_{\rm cloud}$ is a constant specified as a free parameter, i.e., a cloud-to-gas ratio by number.  
We use a constant value, $f_{\rm cloud} = 10^{-17}$. The spatial homogeneity of condensates and the value chosen for $f_{\rm cloud}$ are not meant to correspond to a physically realistic scenario; rather, we are merely choosing this set up to test the code by making scattering and absorption by condensates easily discernible. 

When $k-$tables are used during integration, the cloud properties and the gas scattering cross-section used in Equations \ref{eqn:omega0} and \ref{eqn:g0} are averaged over each $k-$table bin.

\subsubsection{Transition between regular and improved two-stream radiative transfer methods}

In the limit of an opaque (${\cal T}=0$), purely absorbing ($\omega_0=0$), isothermal atmospheric layer, the incoming/outgoing flux becomes $\Pi B$, where $B$ is the Planck function.  The coupling coefficients become $\zeta_+=1$ and $\zeta_-=0$, independent of the value of $g_0$.  However, one obtains $\Pi = \pi / E$.  In this limit, one should recover $E=1$; note that, in the two-stream approach, an atmosphere with $g_0=1$ behaves like a purely absorbing one \citep{heng14}.  Therefore, we expect $E \rightarrow 1$ and $g_0 \rightarrow 1$ as $\omega_0 \rightarrow 0$.  Equation (31) of \cite{heng18} is consistent with this asymptotic limit (and was calibrated on calculations with $\omega_0>0.0025$), but there is no theory on how to approach it.  In the absence of such a theory, we generalise equation (31) of \cite{heng18} to
\begin{equation}
E =
\begin{cases}
\begin{split}
&1.225 - 0.1582 g_0 - 0.1777 \omega_0 - 0.07465 g_0^2 & \omega_0>0.1, \\
&+ 0.2351 \omega_0 g_0 - 0.05582 \omega_0^2, \\
& 1, & \omega_0 \le 0.1. \\
\end{split}
\end{cases}
\end{equation}
The improved two-stream approach is switched off when $\omega_0 \le 0.1$ as it produces similar outcomes to the regular two-stream approach when $\omega_0=0.1$ \citep{hk17}. This procedure ensures that $\pi B$ (with the appropriate correction term for non-isothermal layers) of flux is emitted by each atmospheric layer when it becomes opaque and purely absorbing.

\subsubsection{Operational procedure}
\label{subsect:opproc}

For each \texttt{THOR+HELIOS} GCM run, we execute the following:
\begin{itemize}

\item Each simulation is initialized with a temperature structure given by Guillot profiles. Specifically, we use Equation 27 of \cite{guillot10} with the added collision-induced absorption approximation of \cite{heng11b}, and $T_{\text{irr}} = 2440$ K, $\mu_\star = 0.5$, $\gamma = 0.5$, and $T_{\text{int}} = 100$ K. The high irradiation temperature, $T_{\text{irr}}$, is chosen to produce a temperature in the deep region of $\sim 2200$ K, which starts the model closer to radiative equilibrium.


\item Each GCM run is performed for 3000 Earth days (with each day having 86,400 seconds), which corresponds to 864,000 time steps.  A constant time step of 300 seconds is used.  Monitoring of the global quantities suggest that dynamical equilibrium is attained only after about 1000 days (see Appendix \ref{append:conservation}).  We discard output from the first 2000 days and base our analysis only on output from between Days 2001 to 3000. Note that radiative equilibrium is achieved only for the cloud-free simulation (lower left panel of Figure \ref{fig:global_quantities}). The deep regions are still slowly adjusting at 3000 days in the cloudy cases, though the non-isothermal simulations are converging faster than the isothermal, similar to the convergence issues noted in \cite{malik17} for the 1-D model. Nevertheless, in all simulations, the flow and temperatures do not change noticeably after $\sim 1000$ days.


\item After 3000 days, the GCM is run for one more time step but using the high-resolution opacity file (Figure \ref{fig:opacity}), which has a spectral resolution of 500.  This is a post-processing step to generate synthetic spectra of a higher resolution. For post-processing, we extend the top altitude of the simulation, extrapolating the temperatures down to pressures of $\leq 1$ $\mu$bar. As only the radiative-transfer is run during this step, the instability in the dynamical core is avoided (see Section \ref{subsect:RTcomp}). We assume that the temperatures of each column are isothermal above the original model top and take on the value of the top-most layer. This extrapolation is used in all spectra presented in Section \ref{sect:spectra} and is included as an input option of our post-processing code. Users of the code can specify the desired lowest pressure level of the extrapolation. 

\end{itemize}

For each simulation, we checked that the stellar beam is attenuated well before the bottom of the simulation domain (Figure \ref{fig:beam}).  If insufficient opacity is assumed in the visible/optical range of wavelengths, it is possible for starlight to pass through the entire model atmosphere, hit the bottom of the simulation domain, be artificially reflected upwards and emerge as the synthetic spectrum (not shown).

All simulations include 4th-order horizontal hyperdiffusion and 3-D divergence damping. The dimensionless diffusion coefficients are $D_{\rm hyp} = D_{\rm div} = 0.015$ (see Equation 49 of \cite{mendonca16} and Equations 59 and 60 of \cite{deitrick20}). We further include a 6th-order vertical hyperdiffusion term \citep[see][]{ts04}, with a coefficient $D_{\rm ver} = 0.00375$, which helps reduce noise at the vertical grid level. In order to approximate the breaking of waves in the upper atmosphere and prevent spurious wave reflection, we include a sponge layer in the form of Rayleigh drag \citep{mendonca18b,deitrick20} in the top 25\% of the model domain. Winds and temperatures are damped toward the zonal mean in this region with a minimum time-scale of 1000 seconds. The strength of the sponge is gradually increased from zero at 75\% of the top boundary to full strength at the very top. 

\subsubsection{\texttt{HELIOS} integration and code optimisation highlights}

A major design bottleneck was how to interface \texttt{THOR} and \texttt{HELIOS}, especially since they are largely written in different programming languages. As already mentioned, we rewrote \texttt{HELIOS} in the \texttt{C++} programming language (named \texttt{Alfrodull} for book-keeping reasons) in order to optimize the interfacing with \texttt{THOR}.  Most of the computational cost associated with \texttt{HELIOS} is in solving for radiative-convective equilibrium via iteration \citep{malik17,malik19}.  Since \texttt{THOR}  has its own representation of circulation, including convection \citep{mendonca16,mendonca18b,deitrick20}, this feature of \texttt{HELIOS} is superfluous. There is also no requirement to solve for radiative equilibrium in one dimension, since a more general equilibrium in three dimensions is being solved for via the coupled fluid equations (see Appendix \ref{append:review} or, e.g., Chapter 9 of \citealt{heng17}).  In \texttt{THOR+HELIOS}, the main task of \texttt{Alfrodull} is to transform abundance-weighted, temperature- and pressure-dependent opacities into fluxes, which are then integrated over wavelength to obtain heating and cooling rates.

Firstly, the workflow management of \texttt{HELIOS} was translated to \texttt{C++}.  The code was embedded within a small library that could be used within \texttt{Alfrodull}.  It was verified that the \texttt{C++} translation reproduced the initial algorithm.  Secondly, \texttt{Alfrodull} was interfaced as a physics module to the \texttt{THOR} code, which allowed the former to use the data storage infrastructure of the latter and to access its data.  The data from the vertical spatial grid of \texttt{THOR} are converted to the pressure grid of \texttt{Alfrodull}.

Thirdly, we replace the iterative approach used in \texttt{HELIOS} \citep{malik17,malik19} with the implementation of Thomas's algorithm as described in Section \ref{subsect:thomas}.  This upgrade was motivated by tests suggesting that the implementation of an iterative, pair-wise approach over all of the columns of \texttt{THOR} in three dimensions is computationally prohibitive in practice, as each time step took several minutes to complete.  Thomas's algorithm requires one downward pass to compute the coefficients and one upward pass of back-substitution.  This is roughly equivalent to the iterative approach taking one upward and one downward pass, which implies a gain in computational speed of a factor of roughly ${\cal N}$.  We optimised the algorithm to run in parallel over multiple columns, provided sufficient GPU cores and memory were available.  In practice, successive batches of columns are computed in serial due to memory constraints; within each batch, the columns are computed in parallel.

\section{Benchmarking against 1-D \texttt{HELIOS}}
\label{sect:benchmark}

Here, we run several tests to validate the \texttt{THOR}-coupled radiative transfer by comparing to the standalone 1-D \texttt{HELIOS} code \citep{malik17,malik19}. For this, we run \texttt{THOR+HELIOS} in ``1-D mode'' by switching off the dynamical core and reducing the grid to a single column. The only physical processes acting on the column are the radiative-transfer followed by an adjustment to the pressure and density in each layer to preserve hydrostatic balance. Note that when the dynamical core is enabled, hydrostatic balance is continually sought by the algorithm solving the Euler equations; without the dynamical core, and because \texttt{THOR} utilizes an altitude grid rather than pressure, another mechanism must be enabled to retain hydrostatic balance. This extra step is unnecessary in models that utilize a pressure grid, such as 1-D \texttt{HELIOS}, because hydrostatic balance is usually implicit in the equations.

More concretely, hydrostatic balance is enforced by the following algorithm. After the radiative-transfer module has computed the temperature of each layer, we compute the pressure. The pressure in the lowest layer is held at the reference pressure, $P_{\rm ref}$. The pressure in each layer. $i$, above is set based on the layer below, $i-1$, according to
\begin{equation}
    P_i = P_{i-1}\frac{ \frac{1}{z_i-z_{i-1}}-\frac{g}{2 R_d T_{i-1}}}{\frac{1}{z_i-z_{i-1}}-\frac{g}{2 R_d T_{i}}},
\end{equation}
which is derived from the discrete equation for hydrostatic balance and the ideal gas law. After determining the pressure in each layer, the density is calculated from the ideal gas law. This does not conserve mass as the dynamical core does, but here we are only interested in reaching radiative equilibrium.

We run \texttt{THOR+HELIOS} in this way with 4 different zenith angles assigned to the direct beam: $\theta_{\star} = 0^{\circ}, 30^{\circ}, 60^{\circ},$ and $89^{\circ}$. Each case is run for a total of 800 days, which is more than enough to reach a steady state. We then compare to 1-D \texttt{HELIOS} run under identical conditions. The resulting temperature-pressure profiles are plotted in the left panel of Figure \ref{fig:helios_thor_comp}. For all $\theta_{\star}$ except $89^{\circ}$, the profiles are nearly identical. For $\theta_{\circ} = 89^{\circ}$, there are minor differences but the models still match reasonably well. 

We summarize the differences between the two models (1-D \texttt{HELIOS} and \texttt{THOR+HELIOS} run in 1-D mode) here:
\begin{itemize}
    \item 1-D \texttt{HELIOS} utilizes a pressure coordinate, while \texttt{THOR+HELIOS} uses an altitude coordinate. 
    \item Hydrostatic equilibrium is implicit in the use of pressure in the equations in \texttt{HELIOS}, though the assumption is relevant only for the calculation of layer heights. In \texttt{THOR+HELIOS}, hydrostatic equilibrium is not assumed in the radiative transfer equations, and therefore we adjust the density at the end of each step to restore balance.
    \item The number of layers is 105 in \texttt{HELIOS} and 40 in \texttt{THOR+HELIOS}. The number of layers in the latter is chosen to be the same as in the full 3-D simulations. 
    \item \texttt{THOR+HELIOS} uses a real heat capacity and a physical time-step, while in \texttt{HELIOS} the heat capacity is ignored and the time-step adjusts based on the heating rates. 
    \item \texttt{HELIOS} runs until radiative equilibrium is achieved, i.e., until the upward and downward fluxes (or equivalently, the net fluxes) at each layer interface approach a constant value within some tolerance. \texttt{THOR+HELIOS} does not check for radiative-equilibrium and so we simply run this model until we observe a steady state. 
\end{itemize}

\begin{figure*}
\begin{center}
\vspace{-0.1in}
\includegraphics[width=\textwidth]{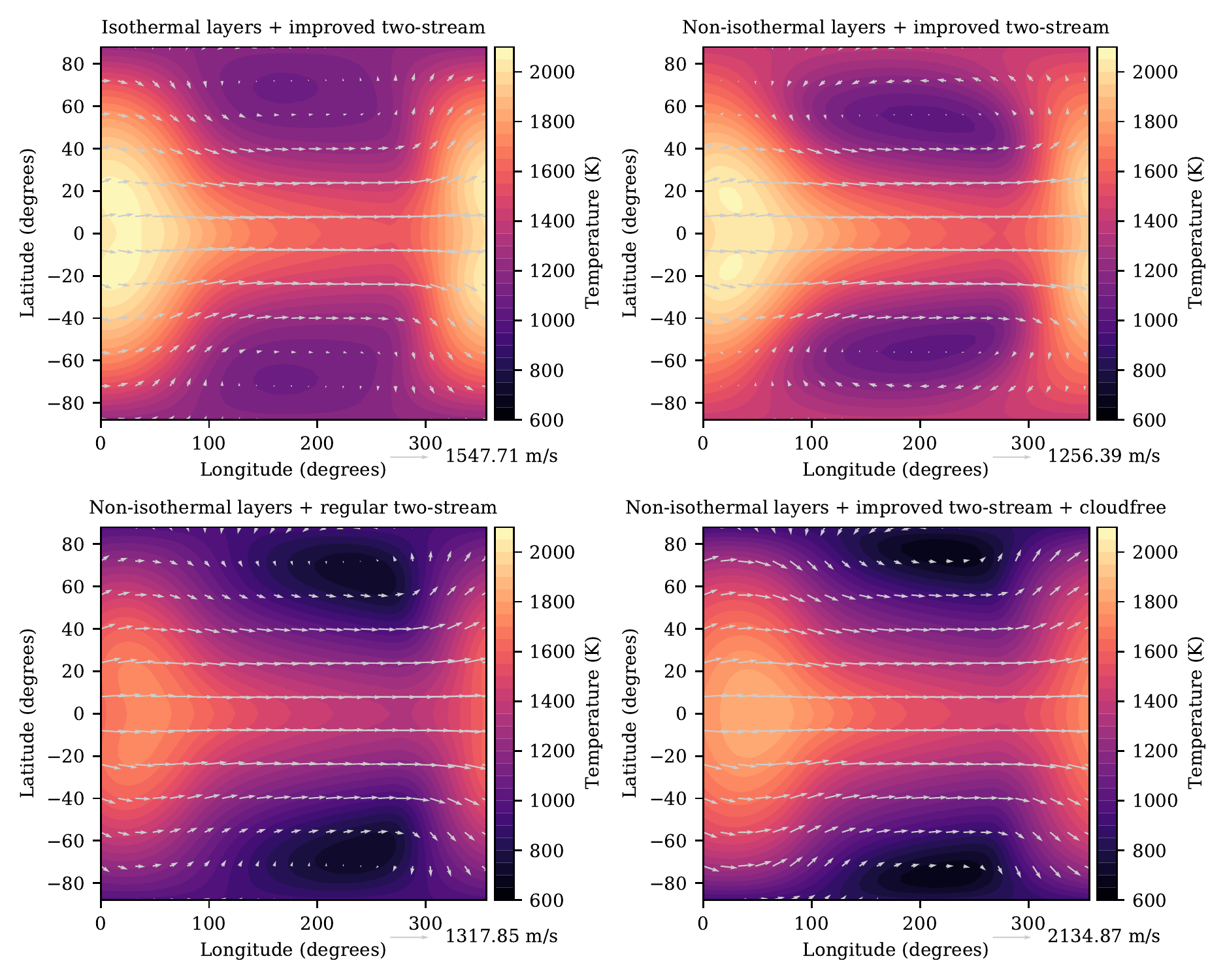}
\end{center}
\caption{ Temperature slice of each of the four GCMs presented in the current study, as labelled in each subpanel. The selected altitude is  1200 km from the bottom of the simulation domain, which corresponds to $\sim 0.1$ bar. The scale of wind vectors is indicated in the lower right of each panel.}
\vspace{0.1in}
\label{fig:potential_temp_slices}
\end{figure*}

Figure \ref{fig:helios_thor_comp} also shows the contribution function calculated from 1-D \texttt{HELIOS}. 
In \texttt{THOR} hot Jupiter simulations, because of altitude coordinate and the strong day-night dichotomy, the pressure at the top of the model on the day-side of the planet can be 3--4 orders of magnitude higher than the pressure at the top on the night-side. In our 3-D WASP-43b simulations (Section \ref{sect:results}), we reach pressures of $\sim 10^{-3}$ bar on the day-side and $\sim 10^{-7}$ on the night-side. As we see in Section \ref{subsect:RTcomp}, capturing the complete contribution function is a challenge in the full 3-D simulations. 

To verify that our spectral resolution is sufficient, we run 1-D \texttt{HELIOS} with several different resolutions and two different sampling methods, k-distributions and opacity sampling. The resulting temperature-pressure profiles are shown in Figure \ref{fig:helios_specres_test}. Each T-P profile is compared to the 300 bin k-table simulation (which contains the greatest amount of spectral information) in the right panel of Figure \ref{fig:helios_specres_test}. While large errors occur for the 10 bin k-table simulation and the $R=50$ opacity sampling solution, the others compare very well. We run our 3-D simulations with the 30 bin k-table, as this provides the optimal balance between computational efficiency and numerical accuracy. With current GPUs and under our current set-up, running 3-D simulations with 100 or 300 bin k-tables is computationally unfeasible. Additionally, opacity sampling at $R=50$ is nearly as computationally expensive as using 30 k-table bins, and less spectral information is resolved. Future model design improvements, such as expanding the code to multiple GPUs, should make it possible to integrate with higher spectral resolution.

\section{Four GCMs of WASP-43b}
\label{sect:results}

To showcase the technical developments made, we construct four GCMs of the hot Jupiter WASP-43b.  Due to its short (<24 hours) orbital period, WASP-43b is one of the few exoplanets for which one may obtain multi-wavelength phase curves using the Wide Field Camera 3 (WFC3) of the Hubble Space Telescope (HST) \citep{stevenson14}.  Several previous studies have presented GCMs of WASP-43b (e.g., \citealt{kataria15,mendonca18a,mendonca18b}).  In the current study, our four GCMs of WASP-43b include:

\begin{figure*}
\begin{center}
\vspace{-0.1in}
\includegraphics[width=\textwidth]{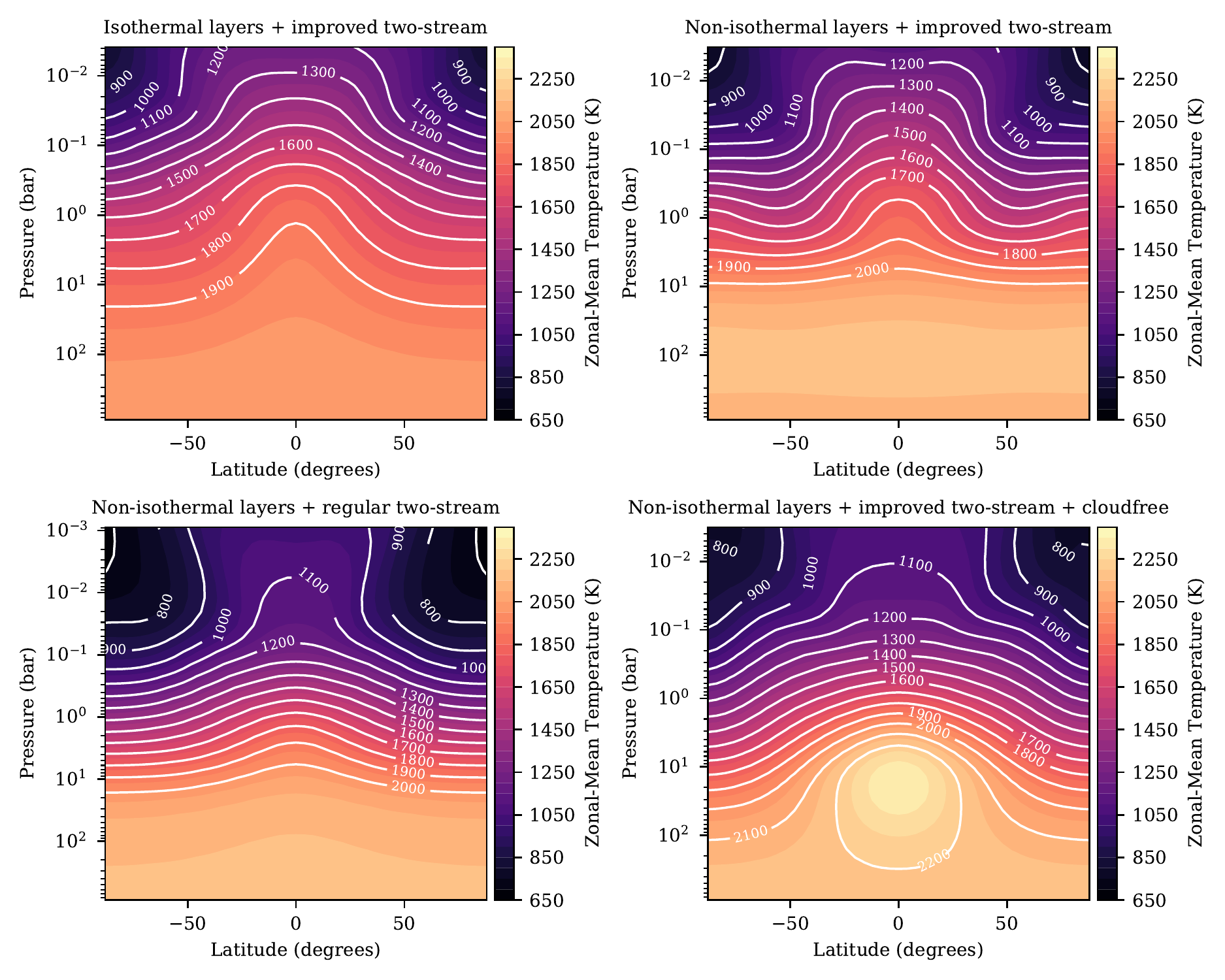}
\end{center}
\vspace{-0.2in}
\caption{Zonal-mean temperature profiles of the four GCMs presented in the current study, as labelled in each subpanel.}
\vspace{-0.1in}
\label{fig:zonal_mean_temperature}
\end{figure*}

\begin{enumerate}

\item A radiative transfer model with isothermal layers (constant Planck function), while implementing the improved two-stream method, with enstatite condensates throughout the atmosphere.

\item Non-isothermal layers (which include the gradient of Planck function) with improved two-stream radiative transfer, with enstatite condensates throughout the atmosphere.

\item Non-isothermal layers with regular, hemispherical two-stream radiative transfer, with enstatite condensates throughout the atmosphere.

\item Non-isothemal layers with improved two-stream radiative transfer, but assuming a cloud-free atmosphere.

\end{enumerate}
As already mentioned, our implementation of clouds in the first three GCMs are for the purpose of studying the effects of scattering as modelled using improved versus regular two-stream radiative transfer, rather than any attempt to be realistic about cloud physics.  The consideration of isothermal layers in the first GCM is motivated by the study of \cite{malik17}, which demonstrated that one-dimensional radiative transfer models struggle to converge to radiative equilibrium even with 1001 isothermal layers, whereas models with 21 non-isothermal layers do attain convergence.

Table \ref{tab:parameters} contains the input parameters of the GCM for WASP-43b, which were curated from the published literature. All simulations include an internal heat flux at the bottom boundary, $F_{\rm BOA}$, with an emission temperature of 100 K.

\subsection{Estimates of computational speed}

We provide two suites of estimates of computational speed.  The first suite focuses solely on the \texttt{THOR} dynamical core: reproducing the Held-Suarez Earth benchmark \citep{hs94}, which does not invoke multi-wavelength radiative transfer.  To match the original \cite{hs94} study, we used a horizontal grid resolution of $g_{\rm level}=5$, which corresponds to about 2 degrees on the sphere.  We used 32 vertical levels.  We ran each simulation for 300 time steps with a time step of 1000 seconds on four different types of GPUs.  For the NVIDIA GeForce GTX 1080 Ti, GeForce RTX 2080 Ti, Tesla P100, and Tesla K20 GPUs, 300 time steps took about 71, 36, 57, and 353 seconds, respectively, which corresponds to about 6.8, 3.5, 5.5, and 34 hours for the full 1200 Earth days of the Held-Suarez benchmark. Simulations such as these require a lower degree of parallelization than multi-wavelength radiative transfer simulations, where parallelization occurs across wavelength as well. The consumer GPU cards (1080 Ti and 2080 Ti) tend to offer similar or slightly better performance than the professional GPU card (P100) with similar compute capability.

\begin{figure*}
\begin{center}
\vspace{-0.1in}
\includegraphics[width=\textwidth]{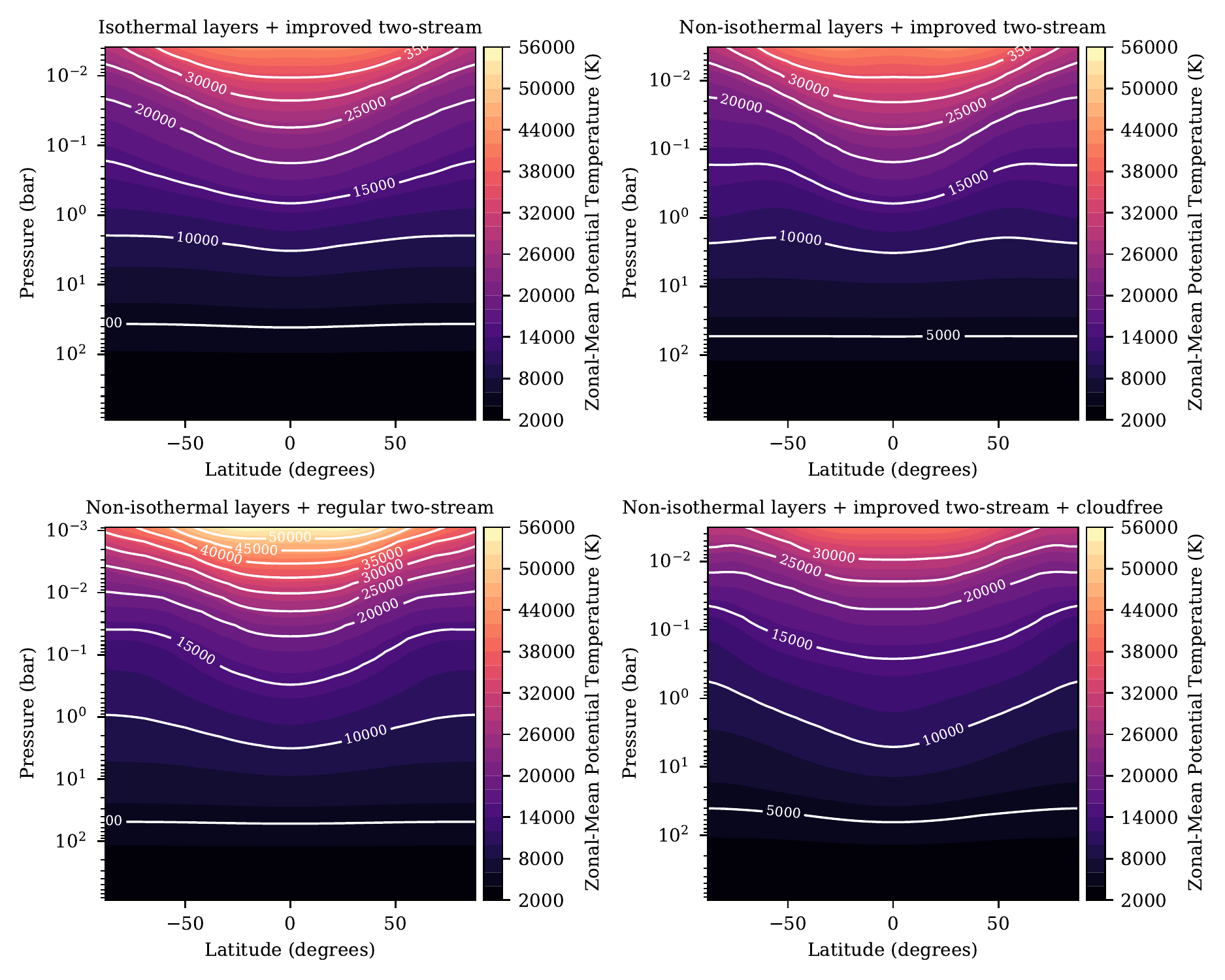}
\end{center}
\vspace{-0.2in}
\caption{Zonal-mean potential-temperature profiles of the four GCMs presented in the current study, as labelled in each subpanel.}
\vspace{-0.1in}
\label{fig:zonal_mean_potential}
\end{figure*}

The second suite of tests focuses on \texttt{THOR+HELIOS} GCMs of WASP-43b and showcases the optimisation efforts achieved in the current study to couple the dynamical core with multi-wavelength radiative transfer. We again ran simulations for 300 time steps. For these simulations, the horizontal resolution used is $g_{\rm level}=4$ (about 4 degrees on the sphere), 40 vertical levels are used and the time step is about 300 seconds.  We use the non-isothermal layer solution. Multi-wavelength radiative transfer makes these simulations significantly more computationally expensive. For the 1080 Ti, 2080 Ti, P100, and K20 GPUs, 300 time steps took about 
 778, 623, 580, and 1366 seconds,
respectively, which by extrapolation correspond to about 
26, 21, 19, and 46 days
for our full 3000-day simulations. The professional P100 GPU card edges out the consumer GPU cards with similar compute capabilities (the 1080 Ti and 2080 Ti).


To compare more directly with the \texttt{THOR+HELIOS} simulations, we also ran the Held-Suarez benchmark at the same resolution ($g_{\rm level} =4$) and number of vertical levels (40) for 1200 steps. For the 1080 Ti, 2080 Ti, P100, and K20 GPUs, these took 70, 47, 63, and 365 seconds, respectively. Since the two types of simulations (Held-Suarez benchmark and WASP-43b) require different time step sizes, we can compare the time required for a fixed number of time steps, which will give an estimate of the additional time required by the coupling to \texttt{HELIOS}. To run 10$^{6}$ time steps, the Held-Suarez benchmark takes roughly 16, 11, 14.5, and 84.5 hours on the 1080 Ti, 2080 Ti, P100, and K20 GPUs, respectively. The  \texttt{THOR+HELIOS} simulations take about 
 720, 577, 537, and 1265 hours
to do 10$^6$ time steps on the same GPUs. The 2080 Ti has more GPU cores than the P100, but the P100 has better double-precision compute power.  However, since we do not see this benefit of the P100 in the Held-Suarez test, we may surmise that the calculation is dominated by memory access.  We conclude that our implementation of \texttt{HELIOS} (\texttt{Alfrodull}) is not memory-access-limited, which enables the P100 GPU to run optimally.

\subsection{Basic climatology}

\begin{figure*}
\begin{center}
\vspace{-0.1in}
\includegraphics[width=\textwidth]{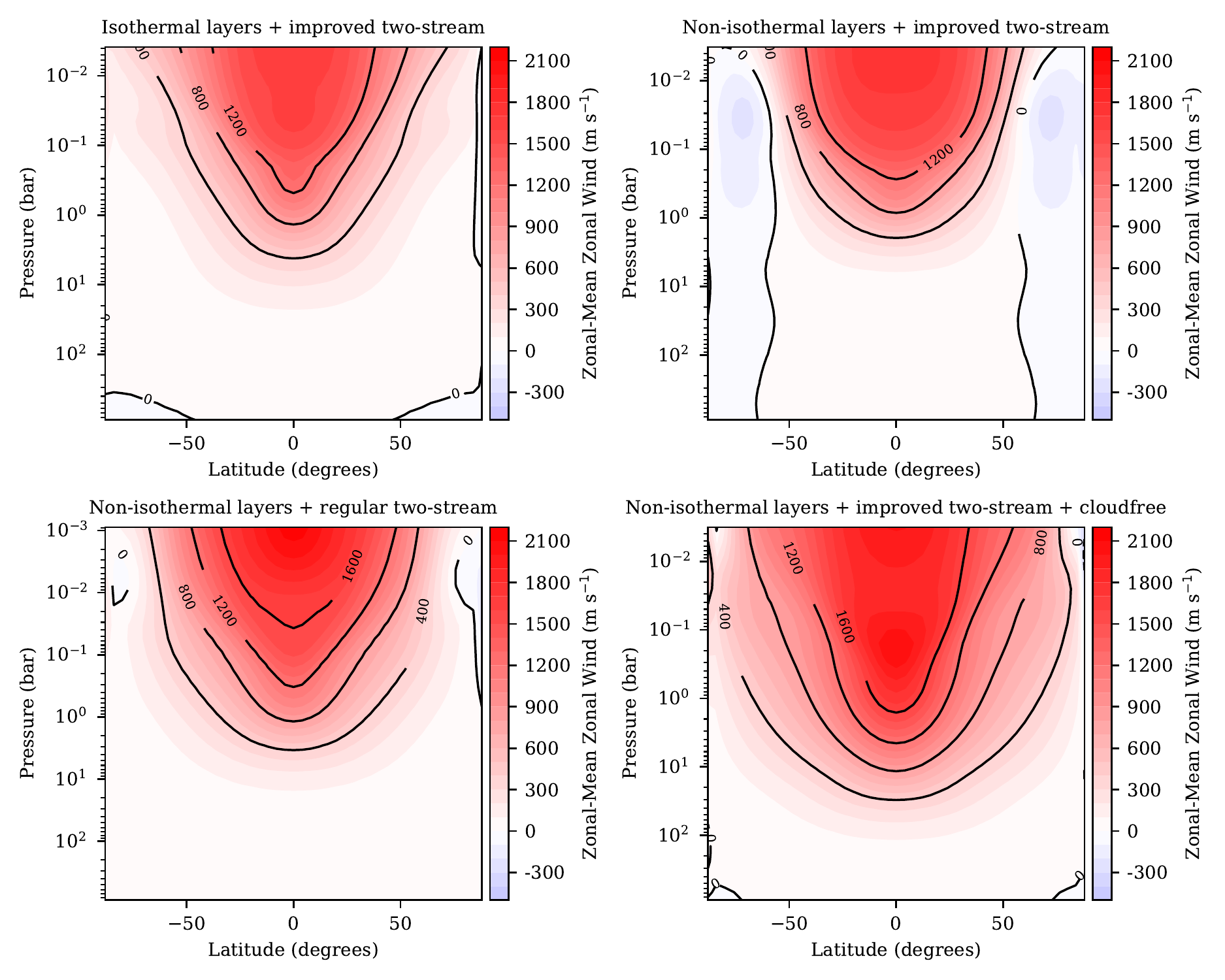}
\end{center}
\vspace{-0.2in}
\caption{Zonal-mean zonal wind profiles of the four GCMs presented in the current study, as labelled in each subpanel.}
\vspace{-0.1in}
\label{fig:zonal_mean_zonal_wind}
\end{figure*}

\begin{figure*}
\begin{center}
\vspace{-0.1in}
\includegraphics[width=\textwidth]{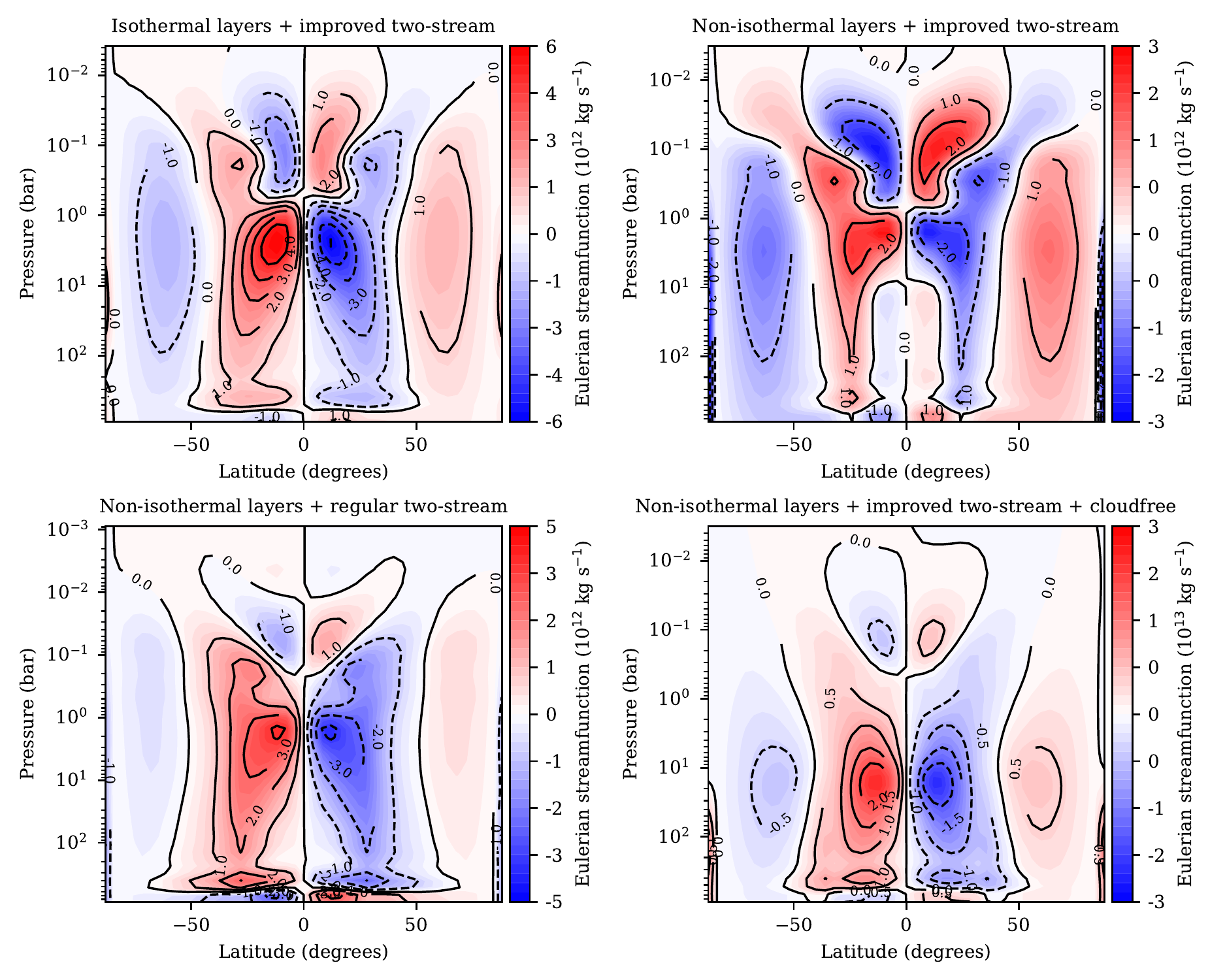}
\end{center}
\vspace{-0.2in}
\caption{Eulerian-mean streamfunction profiles of the four GCMs presented in the current study, as labelled in each subpanel. Note that the scale in the lower right panel is an order of magnitude higher than the other three panels.}
\label{fig:zonal_mean_streamfunction}
\end{figure*}

For the four GCM models presented in this study, the familiar chevron-shaped feature \citep{sp11,tsai14} is shown in Figure \ref{fig:potential_temp_slices}. The zonal-mean profiles are shown in Figure \ref{fig:zonal_mean_temperature} (temperature), Figure \ref{fig:zonal_mean_potential} (potential temperature), Figure \ref{fig:zonal_mean_zonal_wind} (zonal wind), and Figure \ref{fig:zonal_mean_streamfunction} (streamfunction). 
Since a non-hydrostatic GCM is being used, $P$ is not a coordinate like in a GCM that solves the primitive equations, but rather a quantity that varies with time and location.  Therefore, to produce Figures \ref{fig:zonal_mean_temperature} to \ref{fig:zonal_mean_streamfunction} we have computed the temporally, latitudinally (meridionally) and longitudinally (zonally) averaged pressure $\bar{P}$.  The maximum value of this one-dimensional array is $\bar{P}_{\rm max}$.  

The Eulerian-mean streamfunction is defined as \citep{po84,pau08}
\begin{equation}
 \Psi = \frac{2 \pi R_p \cos{\theta}}{g} \int_{0}^{\bar{P}} \bar{v}_\theta ~d\bar{P},
\end{equation}
where $\theta$ is the latitude and $\bar{v}_\theta$ is the temporally and zonally averaged meridional velocity.  The convention is chosen such that positive values of $\Psi$ correspond to clockwise circulation \citep{fhz06}.  Note that the preceding expression is slightly different from equation (21) of \cite{heng11b}, who omitted the factor of $2\pi$.  
 The integral was computed using a trapezoidal rule (the \texttt{trapz} 
routine in Python \citep{vir20}.) 
Figure \ref{fig:zonal_mean_streamfunction} shows the large-scale circulation cells with air \textit{descending} at the equator at pressures above $\sim 0.1$ bar (opposite from the case of Earth), a phenomenon that was previously elucidated by \cite{sp11}, \cite{tsai14}, \textbf{ \cite{charnay15},} and \cite{mendonca20}.  Figure \ref{fig:zonal_mean_streamfunction}, which performs the averaging of the meridional velocity used to construct $\Psi$ over the entire range of longitudes, does not reveal the structure of these circulation cells.

To construct Figure \ref{fig:zonal_mean_potential}, the zonal-mean potential temperature is defined as
\begin{equation}
\bar{\Theta} = \overline{T \left( \frac{{P}}{{P}_{\rm ref}} \right)^{-\kappa}},
\end{equation}
where $\kappa = {\cal R}/c_P \approx 2/7$ is the adiabatic coefficient \citep{pierrehumbert,heng17} and $\bar{T}$ is the zonal-mean temperature. 
These zonal-mean profiles are qualitatively consistent with those reported in previous studies (e.g., \citealt{showman09,heng11a,heng11b,mayne14a,mayne14b,deitrick20}), showing an irradiated atmosphere that is stable against convection and possessing a super-rotating jet at the equator and large-scale circulation cells.  Unsurprisingly, temperatures in the cloudfree GCM are the highest among the four (Figure \ref{fig:zonal_mean_temperature}) as the absence of clouds means that less starlight is reflected away and more heating of the atmosphere occurs.  
The zonal jet in this case is faster and broader than in the cloudy cases (Figure \ref{fig:zonal_mean_zonal_wind}), while the meridional circulation in the deep region is stronger by roughly an order of magnitude (Figure \ref{fig:zonal_mean_streamfunction}). 
The stronger heating leads to a stronger jet (higher velocities; Figure \ref{fig:zonal_mean_zonal_wind}) and more vigorous circulation (Figure \ref{fig:zonal_mean_streamfunction}).


\subsection{Comparison of radiative-transfer solutions}
\label{subsect:RTcomp}

Figures \ref{fig:potential_temp_slices} to \ref{fig:zonal_mean_streamfunction} demonstrate that the qualitative features of the climatology are only mildly sensitive to whether regular versus improved two-stream radiative transfer---or whether isothermal versus non-isothermal layers---is employed. The same general atmospheric structure appears in all cases. Still, some differences are noteworthy. Temperatures are higher near 0.1 bar for the improved two-stream solution (see next paragraph). The improved two-stream solution also has a slower equatorial jet and weaker meridional circulation. The non-isothermal layers solutions also present retrograde flow at high latitudes, a feature that is absent from isothermal solution in its zonal average. 

\begin{figure*}
\begin{center}
\vspace{-0.15in}
\includegraphics[width=0.95\textwidth]{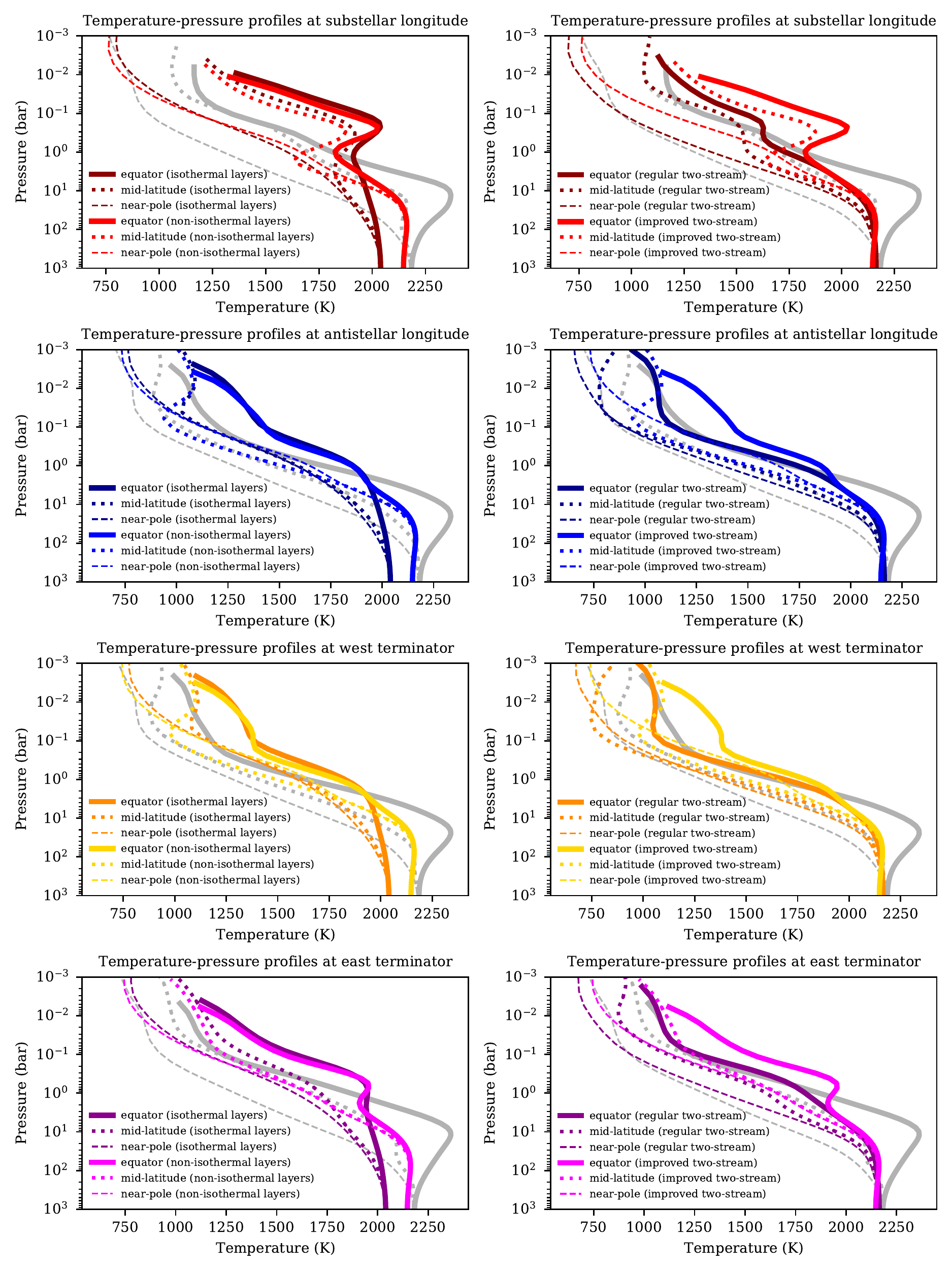}
\end{center}
\vspace{-0.2in}
\caption{Temperature-pressure profiles comparing isothermal versus non-isothermal radiative transfer (left column) and regular versus improved two-stream radiative transfer (right column) at the substellar longitude (first row), antistellar longitude (second row), west terminator longitude (third row) and east terminator longitude (fourth row).  For each longitude, the equator, mid-latitude and near-pole temperature profiles are shown.  For each panel, the gray curves are of the cloudfree case.}
\vspace{-0.1in}
\label{fig:TP}
\end{figure*}

Temperature-pressure profiles at various locations are plotted for all simulations in Figure \ref{fig:TP}. Here, differences in the temperatures around $\sim 0.1$ bar are discernible for the regular and improved two-stream (right column); mainly, the regular two-stream is cooler in this region, consistent with the finding that this method over-estimates back-scattering of photons \citep{kitzmann13,hk17}. 

Though the isothermal layer solution produces qualitatively similar results to the non-isothermal layer solution, we can see a large departure of the temperature in the deep region (Figure \ref{fig:TP}). The isothermal solution is $\sim 200$ K cooler at pressures above $\sim 10$ bar. \cite{malik17} observed the same issue in their simulations with 1-D \texttt{HELIOS} (see their Figure 9), noting that the isothermal layer solution requires $\gtrsim 1000$ vertical layers to achieve convergence, while the non-isothermal merely requires $\sim 20$. Using 1000 or more layers in \texttt{THOR+HELIOS} is computationally infeasible. Moreover, we find that the speed increase afforded by the isothermal solution is less than a factor of 2, compared to the non-isothermal with the same number of vertical levels. For these reasons, we strongly advise the usage of non-isothermal layers.

The temperature-pressure profiles in Figure \ref{fig:TP} reveal a limitation of non-hydrostatic GCMs applied to hot Jupiter forcing regimes---the pressure at the top of the model in the hottest locations is between $10^{-2}$ to $10^{-3}$ bar. This limitation occurs because the dynamical core is unstable below $\sim 10^{-6}$ bar. In the present forcing regime, when the pressure at the top of the model in the coolest locations is $\lesssim 10^{-6}$ bar, the top boundary pressure in the hottest regions is up to 4 orders of magnitude larger. We ran a number of additional simulations that increased the height of the top boundary, but were unable to stabilize any reaching higher than the ones presented here. Note that prior \texttt{THOR} simulations of hot Jupiters \citep{mendonca18a,mendonca18b,deitrick20} have extended to pressures $<10^{-3}$ bar on the day-side, including for WASP-43b, when dual-band gray radiative transfer was used. The important difference is that the multi-wavelength radiative transfer in this case enhances the day-night temperature difference, increasing the probability of triggering the instability. 

\begin{figure*}
\begin{center}
\vspace{-0.15in}
\includegraphics[width=0.95\textwidth]{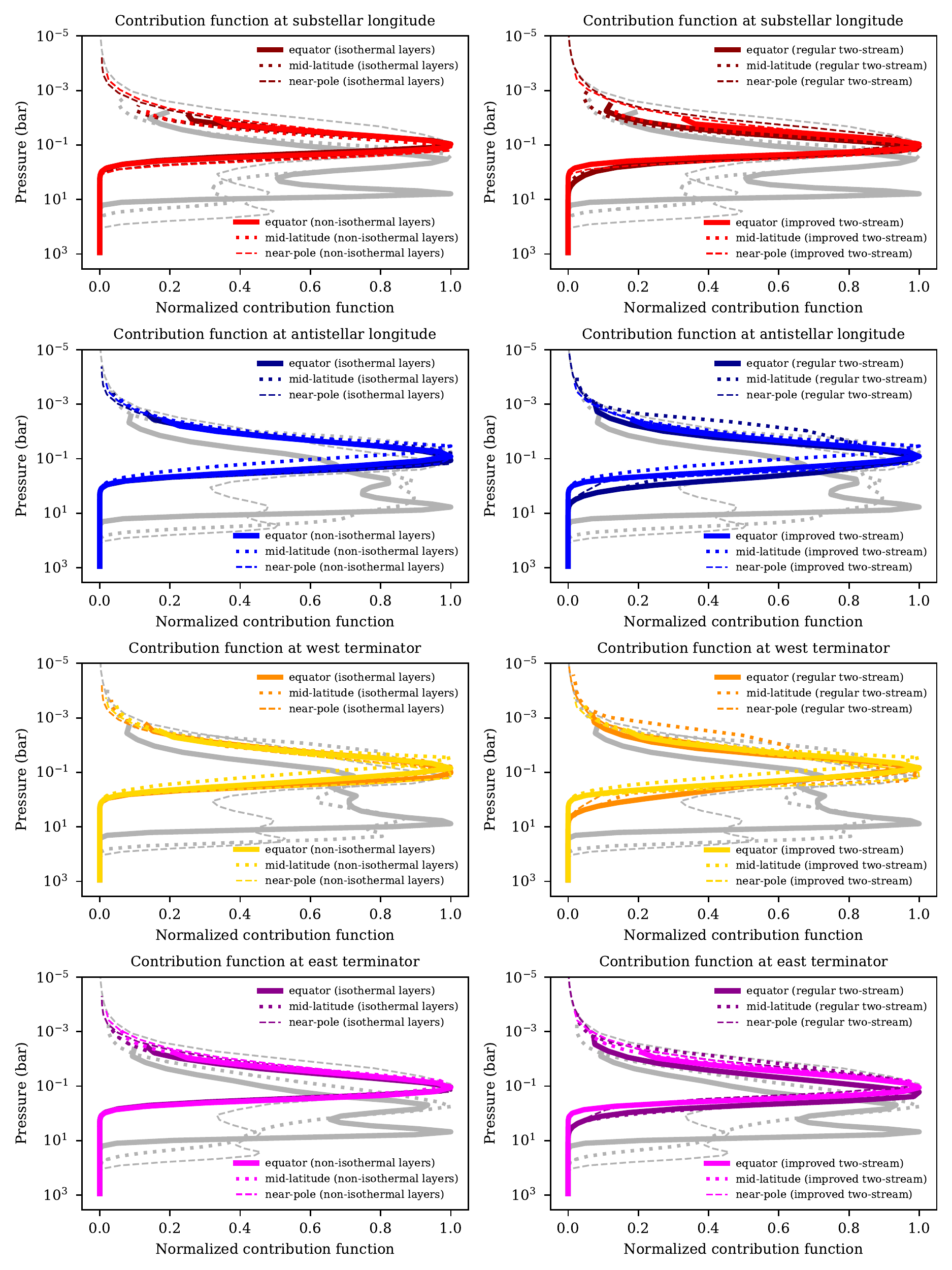}
\end{center}
\vspace{-0.2in}
\caption{ Wavelength-integrated contribution function as a function of pressure, comparing isothermal versus non-isothermal radiative transfer (left column) and regular versus improved two-stream radiative transfer (right column) at the substellar longitude (first row), antistellar longitude (second row), west terminator longitude (third row) and east terminator longitude (fourth row).  For each longitude, the equator, mid-latitude and near-pole temperature profiles are shown.  For each panel, the gray curves are of the cloudfree case. The planet's photosphere is defined by the peak of the contribution function. In the cloud-free case, there is more than one peak---an indication that there are ``spectral windows'' at different wavelengths.}
\vspace{-0.1in}
\label{fig:contr3d}
\end{figure*}

Since our primary concern in this paper is to test the radiative transfer module, the more important question is whether or not these simulations extend high enough to capture the peak thermal emission. Figure \ref{fig:contr3d} shows the wavelength-integrated contribution function \citep[Equation 24 of][]{malik19} for all 4 simulations at different locations. The peak of the contribution is captured at all locations, however, it does not reach zero before the model top at locations along the equator. This is one difficulty with our current algorithm that we have been unable to circumvent with numerical diffusion. We are exploring other solutions at present, however, these require a major reworking of the dynamical core code.



\subsection{Reflected light versus thermal emission: synthetic spectra and phase curves}
\label{sect:spectra}

\begin{figure*}
\begin{center}
\vspace{-0.1in}
\includegraphics[width=\textwidth]{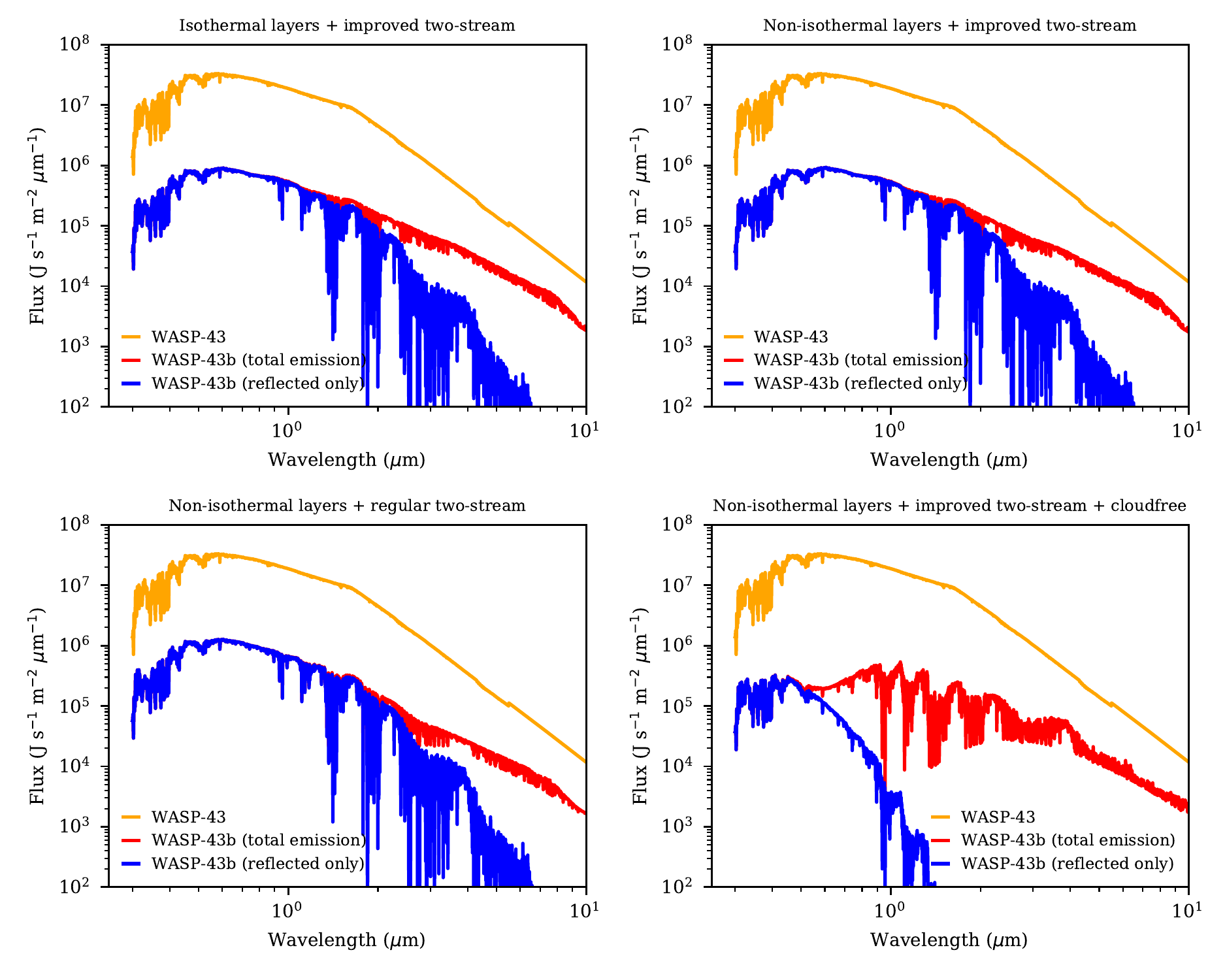}
\end{center}
\vspace{-0.2in}
\caption{Comparing the synthetic spectra of our WASP-43b GCMs with the incident stellar spectrum at the substellar point. Planet spectra are post-processed at the end of each 3000 day simulation, using opacity sampling at $R = 500$.}
\vspace{-0.1in}
\label{fig:spectrum_substellar}
\end{figure*}
\begin{figure*}
\begin{center}
\vspace{-0.1in}
\includegraphics[width=\textwidth]{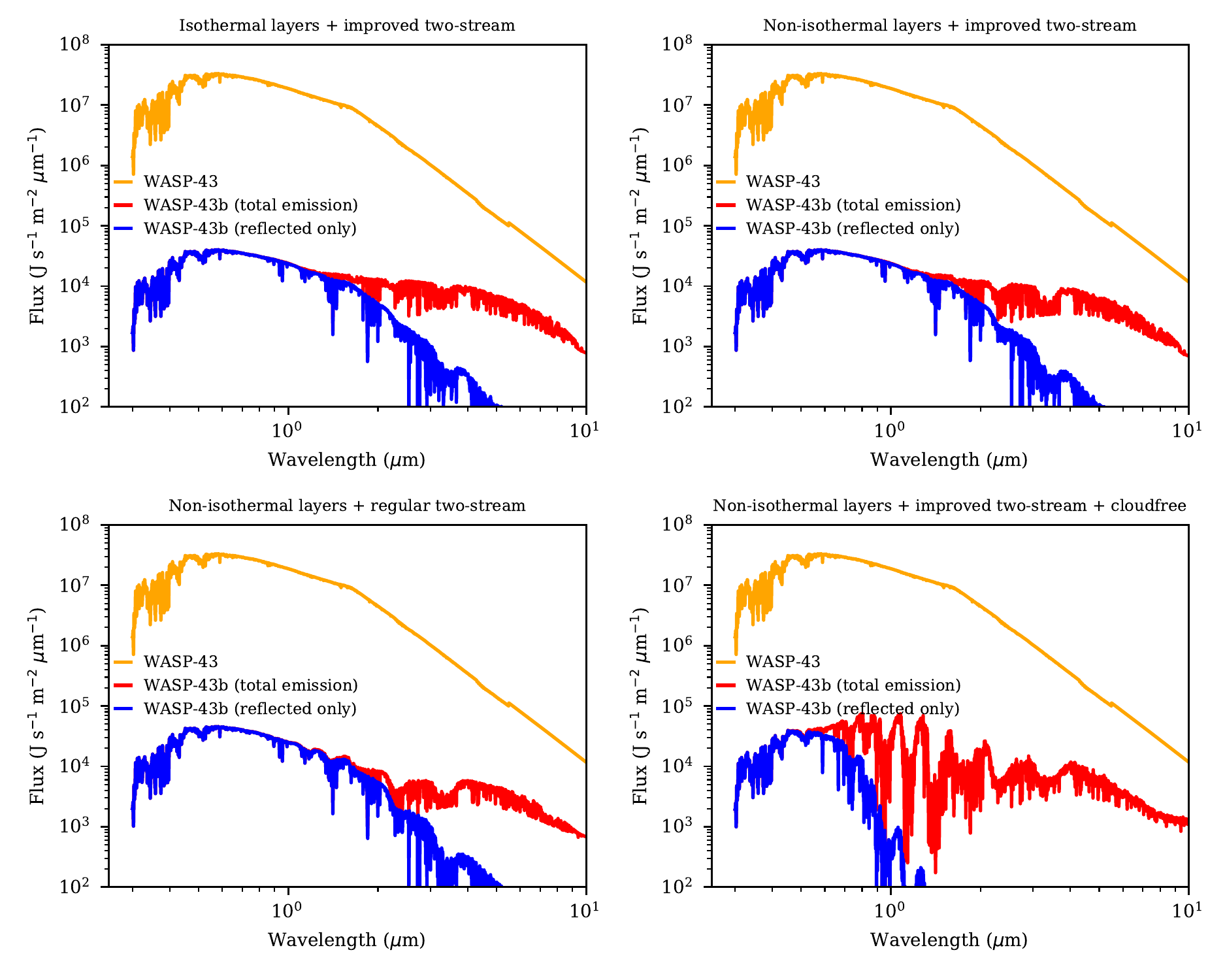}
\end{center}
\vspace{-0.2in}
\caption{Same as Figure \ref{fig:spectrum_substellar}, but at a latitude of 88$^\circ$ (near the north pole).  The longitude is the same as that of the substellar point.}
\vspace{-0.1in}
\label{fig:spectrum_nearpole}
\end{figure*}

 While there are noticeable differences in the global structure of our WASP-43b GCMs, depending on the radiative transfer solution, the output spectra are mostly indistinguishable. Figure \ref{fig:spectrum_substellar} shows the outgoing spectra at the top of the atmosphere at the sub-stellar point. Figure \ref{fig:spectrum_nearpole} shows the same at the substellar longitude near the pole. Other than the cloud-free case, the only discernible differences are in the total emission between 2 and 4 $\mu$m in the polar region. \tempbf{As described in Section \ref{subsect:opproc}, we have extrapolated each column down to a pressure of $\leq 1$ $\mu$bar to produce these spectra and the following phase curves.}

To explore the emission further, we post-process the synthetic spectra at different orbital phases into phase curves by adapting the formalism of \cite{ca08},
\begin{equation}
F = \int^{\lambda_2}_{\lambda_1} \int^{\phi_2}_{\phi_1} \int^{\pi/2}_{-\pi/2} \frac{F_{\rm TOA}}{\pi} ~\cos^2\theta \cos{\left( \phi - \alpha \right)} ~d\theta ~d\phi ~d\lambda, \label{eq:phase}
\end{equation}
where $F_{\rm TOA}$ is the top of atmosphere (TOA) flux, at a given wavelength, emerging from each atmospheric column of the GCM. To simulate the flux measured by the HST-WFC3 instrument, we integrate $F_{\rm TOA}$ from $\lambda_1=1.1$ $\mu$m to $\lambda_2=1.7$ $\mu$m.  Two factors of cosine account for the diminution of flux due to geometric projection across latitude and longitude; the third comes from the solid angle element, $d\Omega$.  The latitude and longitude are denoted by $\theta$ and $\phi$, respectively, while the orbital phase angle is denoted by $\alpha$.  The integration limits, 
\begin{equation}
\phi_1 = -\alpha-\frac{\pi}{2}, ~\phi_2=-\alpha+\frac{\pi}{2},  
\end{equation}
associated with the longitude depend on the exact value of $-\pi \le \alpha \le \pi$ since the GCM adopts the convention of $0 \le \phi \le 2\pi$. 

This integral is most easily performed on the icosahedral grid in the discrete form,
\begin{equation}
     F = \sum_{i}^{n} \frac{F_{\text{TOA},i}}{\pi} \mu_{i} \frac{A_i}{R_p^2},
\end{equation}
where $i$ is the icosahedral grid index, $\mu_i$ is cosine of the angle of each location with respect to the line of sight, and $A_i$ is the area of each control volume at the top of the atmosphere. The solid angle element $d\Omega = \cos{\theta}d\theta d\phi$ becomes $A_i/R_p^2$ in discrete form. By defining $\mu_i$ with the conditions,
\begin{equation}
    \mu_i = 
    \begin{cases}
    \cos{\theta} \cos{(\phi - \alpha)}, & \alpha - \frac{\pi}{2} < \phi < \alpha + \frac{\pi}{2},\\
    0 & \phi > \alpha + \frac{\pi}{2} \text{ or } \phi < \alpha - \frac{\pi}{2}, 
    \end{cases}
\end{equation}
we can do a simple summation over the entire grid to calculate the total received flux. Essentially, this step sets all fluxes emerging from the opposing, out-of-sight hemisphere of the planet to zero.

\begin{figure*}
\begin{center}
\vspace{-0.1in}
\includegraphics[width=\textwidth]{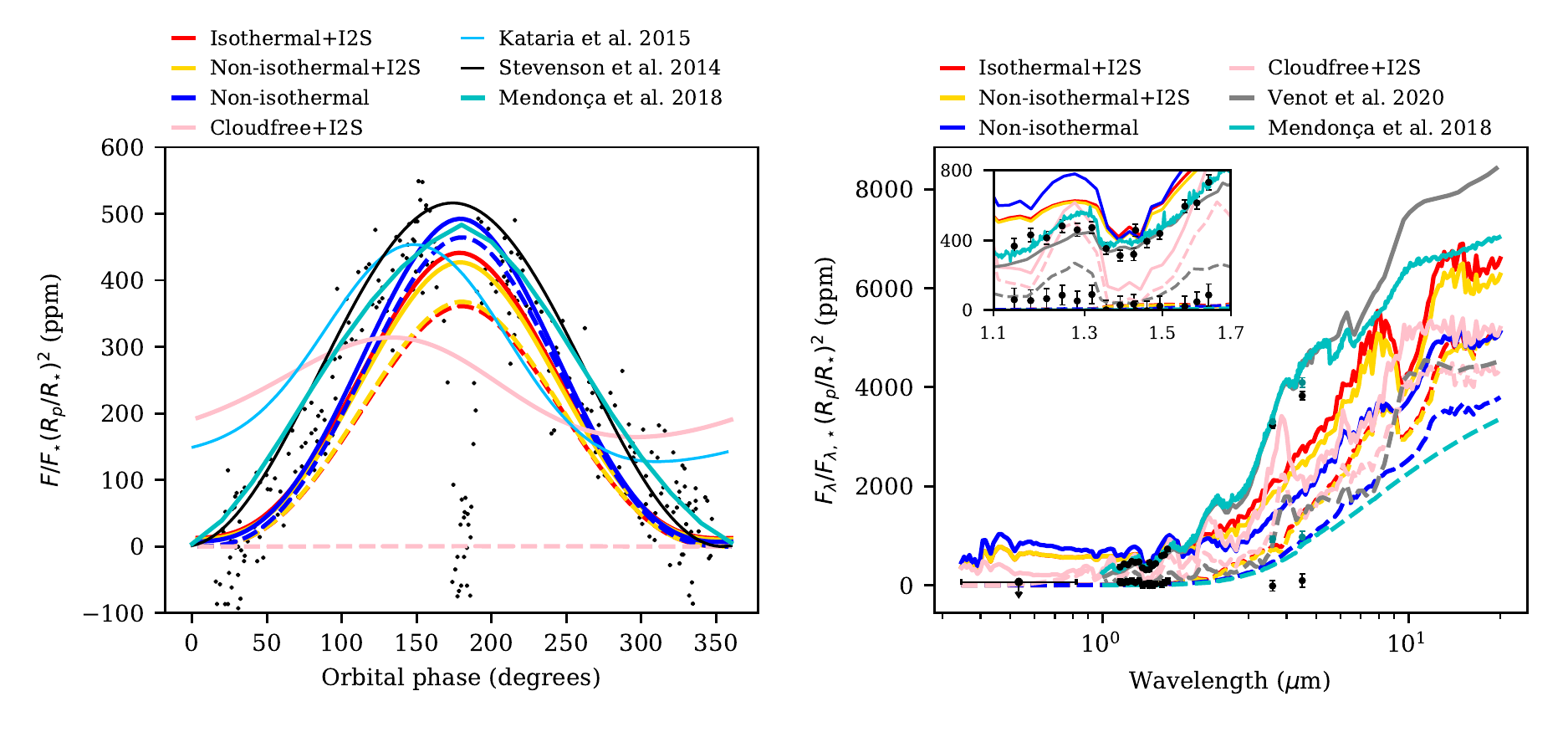}
\end{center}
\vspace{-0.1in}
\caption{\emph{Left:} Phase curves, integrated over 1.1--1.7 $\mu$m to
simulate the HST-WFC3 instrument, of the four WASP-43b GCMs presented in
the current study.  The fluxes have been divided by the stellar flux
integrated over the same range of wavelengths and multiplied by 
$(R_p/R_{\star})^2$ in order to calculate the flux ratio received at 
Earth.  Solid and dashed curves are of the full flux and reflected light 
only, respectively.  ``I2S" is shorthand for ``improved two-stream", while 
``ppm" stands for ``parts per million". The light-blue curve is from 
\protect\cite{kataria15}; it represents their 1$\times$ solar metallicity, cloudy GCM 
simulation without TiO and VO. The cyan curve is from \protect\cite{mendonca18a}; 
a \texttt{THOR} simulation with dual-band gray radiative transfer with 
additional opacity on the night-side to mimic clouds. Black dots are the 
observed, band-integrated values from \protect\cite{stevenson14}; the dip at $\sim 
180^{\circ}$ is the secondary eclipse. The black line is their best fit to 
the phase curve with transit and secondary eclipse omitted 
\protect\citep[supplementary material in][]{stevenson14}.  The data from 
\protect\cite{kataria15} and \protect\cite{stevenson14} was extracted from their plots 
using \texttt{WebPlotDigitizer: https://apps.automeris.io/wpd}. 
\emph{Right:} Dayside (solid) and nightside (dashed) spectra compared to 
the stellar flux for all four models. Model spectra were output at $R=500$ 
and degraded to $R=50$ for plotting. For comparison we include the clear, 
solar composition model from \protect\cite{venot20} in gray \protect\citep[see  also][]{par16,par21}
and the gray radiative transfer case from 
\protect\cite{mendonca18a}, post-processed with 1-D \texttt{HELIOS}. Also included 
are the observations from \protect\cite{stevenson17} as black points and the 
re-processed \textit{Spitzer} points from \protect\cite{mendonca18a} as cyan 
points. \tempbf{The single data point from \protect\cite{fraine21} for wavelengths 346-822 nm 
is an upper limit at 67 ppm.}} 
\vspace{-0.1in}
\label{fig:phase_curves}
\end{figure*}

Figure \ref{fig:phase_curves} shows the HST-WFC3-like phase curves associated with the four WASP-43b GCMs.  The differences between the phase curves employing isothermal versus non-isothermal layers are about 14\% maximum.  However, the difference between the phase curves employing regular versus improved two-stream radiative transfer is about 15\% on average and rises to as high as 38\%. 

When clouds are absent, reflected light is confined to visible/optical wavelengths.  However, when clouds are present, reflected light at $\sim 1$ $\mu$m and longer wavelengths becomes non-negligible (Figures \ref{fig:spectrum_substellar} and \ref{fig:spectrum_nearpole}), a fact further supported theoretical calculations of hot Jupiter albedos \citep[see Figure 13 of][]{morris2021}.  Figure \ref{fig:phase_curves} supports this conclusion, although we note that the eastward peak offset of the phase curve associated with the three cloudy GCMs is about 2$^\circ$ compared to the $12.3 \pm 1^\circ$ shift measured by \cite{stevenson14}.  The cloudfree GCM phase curve has a eastward peak offset of about $45.5^\circ$. Together, these \tempbf{suggest} that the degree of cloudiness present in WASP-43b is non-zero, but less than what we have assumed for the three cloudy GCMs. Nevertheless, these findings suggests that more careful modelling of reflected light versus thermal emission may be required to separate the two components when decontaminating measurements of the geometric albedo via visible/optical secondary eclipse observations.

Figure \ref{fig:phase_curves} also shows the spectra of day-side and night-side for each model, relative to the stellar flux. The cloudy simulations have a much larger difference in the emission between day and night at longer wavelengths than the cloud-free simulation. These simulations also do a bit better at matching the day-side \textit{Spitzer} observations at 3.6 $\mu$m and 4.5 $\mu$m \citep{stevenson17}, though not as well as the cloud-free result from \cite{venot20} or the night-side cloud case from \cite{mendonca18a}. None of the simulations match the original night-side \textit{Spitzer} observations, though we note that \cite{venot20}'s model with enstatite clouds (not shown here) matches these observations well. At the same time, several models are consistent with the re-analyzed night-side \textit{Spitzer} points from \cite{mendonca18a}. In our cloudy models, there is also a small upward tilt in the spectrum toward $\sim 1$ $\mu$m; this is due to scattering by condensates. In fact, it has previously been argued that some reflected light is present in the observations of this planet \citep{keating17}, \tempbf{ though recent measurements suggest the day-side is very dark in the optical and probably cloud free \citep{fraine21}}. The fact that we overestimate the flux at these wavelengths is another indication that the models presented here are too cloudy, particularly on the day-side. \tempbf{At the same time, however, it appears even our cloud-free model produces too much reflection in the optical. Future investigations should attempt to understand this discrepancy.}

\tempbf{We also note that at longer wavelengths (2-20 $\mu$m), our cloud-free GCM produces less flux on the day-side than the similar models from \cite{mendonca18a} and \cite{venot20}. At the same time, the night-side flux in this range is comparable to \cite{venot20} (the simulation from \cite{mendonca18a} is dimmer on the night-side, due to the inclusion of night-side clouds). The list of opacity sources is quite similar between the three simulations, thus the day-side flux should be closer, assuming the same temperatures. In fact, this is the key difference between the simulations---comparing to the simulation from \cite{mendonca18a}, our cloud-free simulation is $\sim500$ K cooler on the day-side around pressures of $0.01-0.1$ bar.  Thus the day-side is fainter in our simulation simply because it is cooler in the photosphere. This is discussed in further detail in Appendix \ref{append:comp2mendonca}. Despite the increase realism of the radiative-transfer solver in this work compared to the \cite{mendonca18a}, the latter of which used dual-band gray radiative-transfer in their GCM, the \cite{mendonca18a} result produces day-side temperatures that better match the \emph{Spitzer} observations. }

\tempbfnew{A potential source of the temperature discrepancy between our cloud-free simulation and that of \cite{venot20} is the lack of alkali species in our opacity list. This was shown to be an important source of opacity by \cite{freedman2008}. Especially their strong, non-Lorentzian line wings can cause significant absorption near the resonance line centers \citep{Allard2016A&A...589A..21A, Allard2019A&A...628A.120A}.
Though this work is primarily focused on detailing our new radiative transfer framework, rather than on explaining all the available data, we acknowledge the importance of species like Na and K and intend to include them in future works.}

For cloud-free models, the \cite{kataria15} simulation is a better fit to the phase curve data than ours in terms of the amplitude and offset. Our cloudy simulations fit the amplitude better, especially regarding the extremely faint night-side. From Figure \ref{fig:contr3d} we can see that the extremely low night-side flux is a consequence of the photosphere being at much lower pressure than in the cloud-free case. However, we see that much of the phase amplitude in these simulations is due to reflection. This and the small phase offset are further indications that we are overestimating the reflected light at short wavelengths ($\sim 1$ $\mu$m). None of the simulations in the present work fits the phase curve quite as well as the prior \texttt{THOR} simulation from \cite{mendonca18a}. This simulation utilized dual-band gray radiative transfer---the spectral information in Figure \ref{fig:phase_curves} was produced via post-processing with 1-D \texttt{HELIOS}.

Future work with \texttt{THOR} will explore this issue for more realistic cloud models, as has been done recently with other GCMs \citep{roman17,roman19,par18,lines19,par21,roman21,christie21}.

Equation \ref{eq:phase} does not take into account limb darkening of the planet, which may be relevant for inflated or low density planets. One method to include limb-darkening may be the use of an empirically tuned model such as used for stars \citep[][for example]{sing09}, applied to the thermal emission. \tempbf{Another method, which is more predictive, is to repeat the two-stream radiative transfer step at output with angle-averaging replaced by the viewing angle of the observer \citep{fortney06}. Yet another predictive method} is to use a Monte Carlo radiative-transfer model for postprocessing, such as \cite{lee17,lee19}, which naturally takes into account the greater path length at the planet limb. It should be noted, however, that symmetric features do not appear in phase curves alone \citep{ca08}, \tempbf{ though spectra may ultimately be affected by the cooler temperatures probed at the limb.} Thus, unless it is strongly asymmetrical, secondary eclipse mapping \citep[e.g.,][]{dewit12} \tempbf{may be} necessary for constraining planetary limb darkening. \tempbf{It is beyond the scope of this paper to include the effects of limb darkening on the emission spectrum, though this is an interesting avenue for future development.}

\section{Discussion}
\label{sect:discussion}

\subsection{Summary of key developments and findings}

In the current study, we report the merging of the \texttt{THOR} GCM and \texttt{HELIOS} radiative transfer solver, as well as the incorporation of improved two-stream (corrected back-scattering) radiative transfer into a GCM. Key aspects of the study include:
\begin{itemize}

\item Radiative transfer is sped up by $\sim 2$ orders of magnitude, compared to the iterative method originally used in the standalone \texttt{HELIOS} code, by implementing Thomas's algorithm to compute multiple scattering of radiation across all layers simultaneously. 

\item Since radiative transfer is performed independently in each atmospheric column, we have invested effort into optimizing it by performing these computations in parallel on a GPU.

\item Using the hot Jupiter WASP-43b as a case study, we show that the global climate is qualitatively robust to whether regular versus improved two-stream radiative transfer or isothermal versus non-isothermal layers are employed, but simulations differ in the finer details. Emission spectra are nearly indistinguishable by eye, however, when integrated to produce HST-WFC3 phase curves, the differences are $\sim 10\%$.

\item The crude assumption of a constant condensate abundance throughout the atmosphere overproduces reflection, as shown by the phase offset and day-side spectrum at $\sim 1.1$ $\mu$m (Fig. \ref{fig:phase_curves}). Nevertheless, the fact that these cloudy simulations match the phase amplitude and offset better than cloud-free simulations indicates that some amount of reflection by clouds is present in the observations. \tempbf{This appears to be contradicted by the extremely shallow secondary eclipse observed by \cite{fraine21}, however, which found that the day-side is very dark and, in all likelihood, cloud-free. Future investigation with a realistic assumption for cloud distribution using \texttt{THOR+HELIOS} should address this contradiction.}

\item A WASP-43b GCM executed for 3000 Earth days with a constant time step of 300 seconds takes approximately 19 days to complete on a Tesla P100 GPU card. The computational time taken for other GPU cards are also reported.

\item The results in Figures \ref{fig:TP} and \ref{fig:contr3d} highlight a challenge of non-hydrostatic modeling of hot Jupiters: instabilities at low pressure prevent us from extending the atmosphere on the day-side to completely capture all components of the radiation. Hydrostatic models, which generally utilize a pressure grid, are capable of reaching pressures of $\sim 1$ $\mu$bar in hot regions, but will become inaccurate when the model domain is $\sim 20\%$ of the planet radius, as is the case for many smaller exoplanets \citep{mayne18}. We are currently exploring solutions to the low pressure instability in \texttt{THOR} and hope to resolve this issue in a future work. \tempbf{Unfortunately, the problem worsens in higher temperature regimes, due to the increasing day-night dichotomy. This prevents us from modeling ultra-hot Jupiters, for example, at present. Users are advised to plot the contributions functions, as we have in Figure \ref{fig:contr3d}, to verify that the bulk of the radiative energy budget is captured by the model domain. }

\end{itemize}



\subsection{Future work}

Future work should replace the simplistic cloud model employed in the current study with a more realistic, first-principles cloud model (e.g., \citealt{lee16,lee17}), where $f_{\rm cloud}$ is a function of location, pressure and temperature \citep{lines19, christie21}.  Ideally, clouds can be modelled using dynamical tracers, although even a static parameterization based on local quantities would be a step beyond our crude assumption in this work.  Chemical disequilibrium driven by atmospheric circulation may be modelled using the technique of chemical relaxation with passive tracers \citep{cs06,drummond18,mendonca18b,tsai18}.  \texttt{THOR+HELIOS} may also be used to simulate ultra-hot Jupiters, but this requires the incorporation of a non-constant specific heat capacity as the atmosphere transitions from being dominated by atomic hydrogen on the dayside to being dominated by molecular hydrogen on the nightside \citep{bc18,par18,komacek18,tan19}.  As we head into the era of JWST, \texttt{THOR+HELIOS} may be used to provide ``null hypothesis" models that assume the same elemental abundances as the host or parent star, where transmission spectra, emission spectra, multi-wavelength phase curves and predictions on variability may be self-consistently computed and confronted by data.

\vspace{0.1in}
\noindent
\textbf{ACKNOWLEDGEMENTS}

\vspace{0.1in}
\noindent
\textit{This paper is dedicated to the memory of Adam P. Showman (1968--2020), a pioneer and global leader in exoplanet GCMs and the study of atmospheric dynamics who had a rare combination of scientific vision and emotional generosity towards newcomers to the field (including KH). We would like to thank both anonymous reviewers for their keen observations and constructive feedback. We would also like to thank O. Venot and V. Parmentier for sharing data used in Figure \ref{fig:phase_curves}. We acknowledge partial financial support from the Center for Space and Habitability (CSH), the PlanetS National Center of Competence in Research (NCCR), the Swiss National Science Foundation, the MERAC Foundation and an European Research Council (ERC) Consolidator Grant awarded to KH (number 771620).  KH acknowledges a honorary professorship from the Department of Physics at the University of Warwick. Calculations were performed on UBELIX (\texttt{http://www.id.unibe.ch/hpc}), the HPC cluster at the University of Bern.  All of the main computer codes used are publicly available as part of the Exoclimes Simulation Platform (\texttt{https://github.com/exoclime}).}

\vspace{0.2in}
\noindent
\textbf{DATA AVAILABILITY}

\vspace{0.1in}
\noindent
\textit{The datasets were derived from sources in the public domain: \texttt{https://github.com/exoclime}}

\clearpage

\bibliographystyle{mnras}
\bibliography{THOR+HELIOS_v3}

\appendix

\section{Review of governing equations of atmospheric dynamics}
\label{append:review}

The Navier-Stokes equation is a mathematical statement of the conservation of momentum for an atmosphere approximated as a fluid (e.g., \citealt{vallis}),
\begin{equation}
\frac{\partial \vec{v}}{\partial t} + \vec{v}.\nabla \vec{v} = \vec{g} - \frac{\nabla P}{\rho} - 2\vec{\Omega} \times \vec{v} + \nu \nabla^2\vec{v} + \frac{\nu}{3} \nabla \left( \nabla . \vec{v} \right),
\label{eq:navier}
\end{equation}
where $\vec{v}$ is the velocity, $t$ is the time, $\vec{g}$ is the acceleration due to gravity, $P$ is the pressure, $\rho$ is the mass density, $\vec{\Omega}$ is the angular rotational frequency of the exoplanet and $\nu$ is the molecular (kinematic) viscosity.  The viscous terms (associated with $\nu$; last two terms in preceding equation) are important only for small length scales (i.e., small Reynolds numbers), typically well below the spatial resolution of the simulation grid, and may be neglected for large-scale circulation.  Dropping these terms yields the Euler equations.  \cite{dd13} retained these viscous terms presumably as a proxy for the turbulent viscosity, since turbulence may be approximated as a viscous process.

If one assumes a steady state for the radial component of equation (\ref{eq:navier}) and neglects the advective, Coriolis and viscous terms, then hydrostatic balance obtains,
\begin{equation}
\frac{\partial P}{\partial r} = \rho g.
\label{eq:hydrostatic}
\end{equation}
Hydrostatic balance or equilibrium is \textit{not} the approximation that the atmosphere is motionless in the radial/vertical direction.  Rather, it is that the timescale for the pressure gradient to balance gravity, which is $\sim H/c_s$ (where $H$ is the pressure scale height and $c_s$ is the sound speed), is the shortest timescale of the system.  Hydrostatic balance occurs essentially instantaneously. In other words, sound waves travel much faster than other waves (e.g., gravity, Rossby) in the system.  In practice, if an explicit integration scheme is used for the radial momentum equation, then the time step of integration is dominated by sound waves and the computational time becomes long \citep{dd12,dd13}, because the computational burden is dominated by having to \textit{solve} for hydrostatic balance.  Assuming hydrostatic balance makes the \textit{choice} of filtering out \textit{all} sound waves at the equation level. One may also implement an implicit integration scheme to selectively filter out sound waves; the HEVI scheme used in the \texttt{THOR} GCM is one such scheme \citep{satoh02,satoh03,ts04,satoh08,mendonca16,deitrick20}.

The mass continuity equation is a mathematical statement of the conservation of mass for a fluid,
\begin{equation}
\frac{\partial \rho}{\partial t} + \nabla . \left( \rho \vec{v} \right) = 0.
\end{equation}
When the approximation of hydrostatic balance is made, it is the mass continuity equation that provides the computation of the radial component of the velocity, since it is set to zero in the radial momentum equation in (\ref{eq:hydrostatic}).

The conservation of energy derives from the first law of thermodynamics for a fluid and is expressed as an evolution equation for the pressure, internal energy, kinetic energy, gravitational potential energy or total energy (e.g., Chapter 9.4 of \citealt{heng17}).  It may also be expressed as an evolution equation for the potential temperature, which is the choice made in the \texttt{THOR} GCM \citep{mendonca16,deitrick20}.  The heating term $Q$, which depends on the spatial gradient of the net flux, is present as one of the terms in the energy equation.

The primitive equations of meteorology are a reduced form of the mass continuity, Euler and energy equations, because three assumptions are made: hydrostatic balance, the shallow atmosphere approximation (where $1/r \approx 1/R$) and the so-called ``traditional approximation" (where the Coriolis and metric terms associated with the radial component of the velocity are neglected (e.g., \citealt{vallis}; Section 9.6.2 of \citealt{heng17}).  

\section{Correspondence of radiative transfer equations used in Heng et al. (2018) versus Malik et al. (2019)}
\label{append:equations}

\begin{figure*}
\begin{center}
\vspace{-0.1in}
\includegraphics[width=\textwidth]{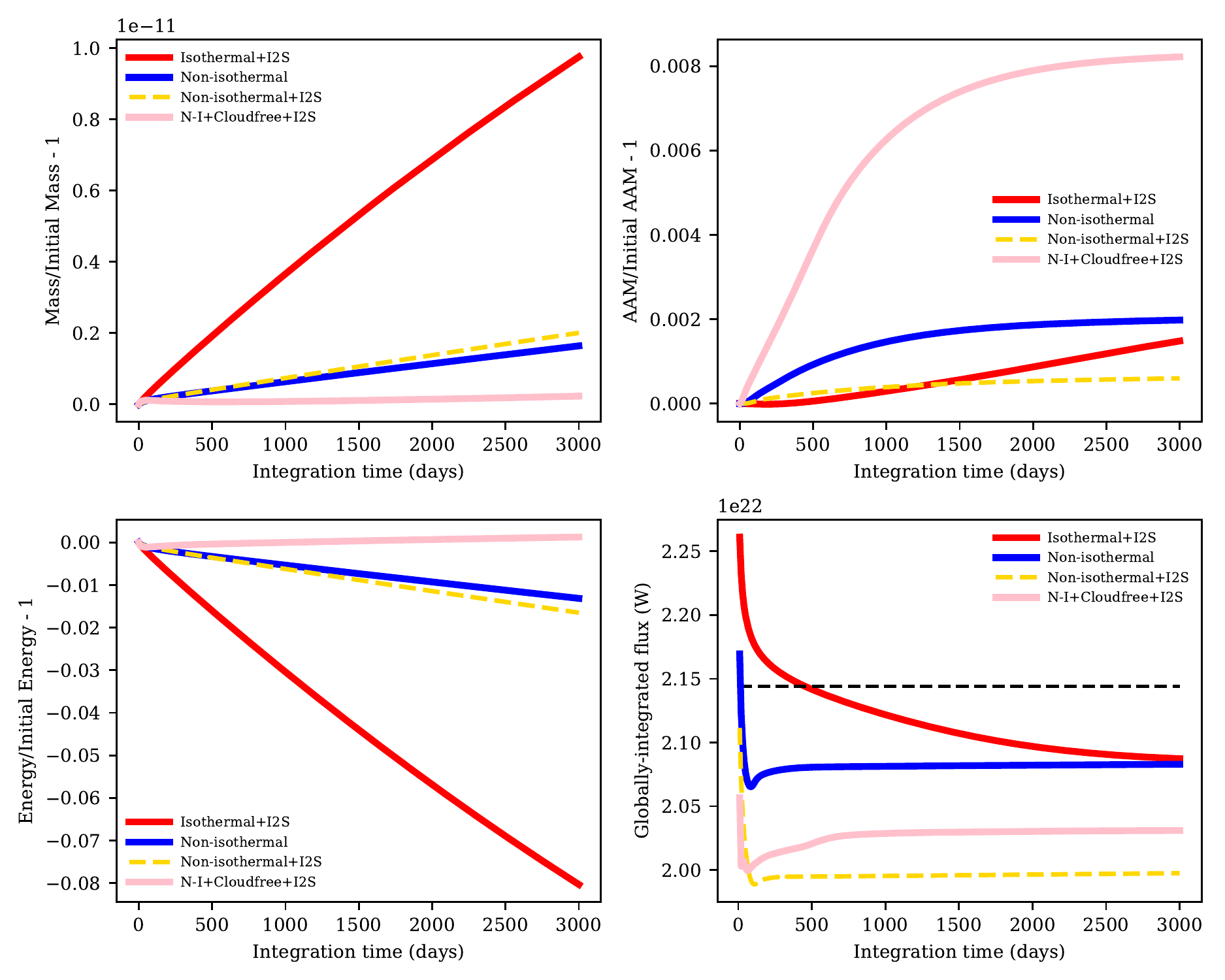}
\end{center}
\vspace{-0.2in}
\caption{Evolution of the global mass (top left panel), axial angular momentum (AAM; top right panel), energy (bottom left panel) and globally-integrated fluxes (bottom right panel) as a function of integration time. See text in Appendix \ref{append:conservation} for description of expectations and precision of the conservation of each quantity. For the lower right panel, the black dashed curve is the incident stellar radiation at the top of the atmosphere; the other curves are the total outgoing fluxes (reflected starlight and thermal emission) at the top of the atmosphere for each model.}
\vspace{-0.1in}
\label{fig:global_quantities}
\end{figure*}

Although \cite{malik19} implement the improved two-stream solutions of radiative transfer in the \texttt{HELIOS} code, equations (8) to (11) are listed for $E=1$ (regular two-stream), where $E$ is the ratio of first Eddington coefficients as defined by \cite{heng18}.  A fitting function for $E(\omega_0, g_0)$ is given in equation (31) of \cite{heng18}. In this section, the equations with $E \ne 1$ (improved two-stream) are cast in the notation of \cite{malik19}, because this corresponds to how they are written in the computer code.  Unlike for \cite{malik19}, the choice of $\epsilon=1/2$ is made, but a value is not chosen for the second Eddington coefficient ($\epsilon_2$).  Following \cite{malik19}, we define
\begin{equation}
\begin{split}
\chi &\equiv \zeta_-^2 {\cal T}^2 - \zeta_+^2, \\
\xi &\equiv \zeta_+ \zeta_- \left( 1 - {\cal T}^2 \right), \\
\psi &\equiv \left( \zeta_-^2 - \zeta_+^2 \right) {\cal T}, \\
\end{split}
\label{eq:coefficients}
\end{equation}
where the coupling coefficients and transmission function are \citep{heng18}
\begin{equation}
\begin{split}
\zeta_\pm &= \frac{1}{2} \left[ 1 \pm \sqrt{\frac{E - \omega_0}{E\left(1-\omega_0 g_0 \right)}} \right], \\
{\cal T} &= e^{-2\sqrt{E\left(E-\omega_0\right)\left(1-\omega_0g_0\right)} ~\Delta \tau},
\end{split}
\end{equation}
and the difference in optical depth between two points is given by $\Delta \tau$.

Consider an atmospheric layer with center index $i-1$ and interfaces indexed by $i-1$ (lower interface) and $i$ (upper interface).  For the outgoing (upward; $F_{\uparrow i}$) and incoming (downward; $F_{\downarrow i-1}$) fluxes, the boundary conditions are, respectively, 
\begin{equation}
\begin{split}
&\frac{1}{\chi}\left( \psi F_{\uparrow i-1} - \xi F_{\downarrow i} \right), \\
&\frac{1}{\chi}\left( \psi F_{\downarrow i} - \xi F_{\uparrow i-1} \right).
\end{split}
\end{equation}
The blackbody terms are also straightforward to write down,
\begin{equation}
\begin{split}
&\frac{\Pi}{\chi} \left[ B_i \left( \chi + \xi \right) - \psi B_{i-1} + \frac{B^\prime}{2E \left( 1 - \omega_0 g_0 \right)} \left( \chi - \psi - \xi \right) \right], \\
&\frac{\Pi}{\chi} \left[ B_{i-1} \left( \chi + \xi \right) - \psi B_i + \frac{B^\prime}{2E \left( 1 - \omega_0 g_0 \right)} \left( \xi + \psi - \chi \right) \right],
\end{split}
\label{eq:bb_terms}
\end{equation}
where we have defined \citep{heng18}
\begin{equation}
\Pi \equiv \pi \frac{1-\omega_0}{E-\omega_0}.
\label{eq:bigpi}
\end{equation}
For the first equation of (\ref{eq:bb_terms}), we note that \cite{malik19} erroneously wrote $\psi B_{i-1}$ as $\xi B_{i-1}$ in their equation (9). We verified that this is a typographical error that does not propagate into the \texttt{HELIOS} code.

For the terms associated with the direct stellar beam, several differences between the notation of \cite{heng18} and \cite{malik19} need to be reconciled.  The beam impinges upon the atmosphere at an angle $\theta_\star$.  Let $\mu_\star \equiv \cos\theta_\star$.  \cite{heng18} defines $\mu_\star$ as a positive quantity, whereas \cite{malik19} defines it as a negative quantity.  This difference in notation causes a flip in sign and implies that $C_\pm / F_\star$ (notation of \citealt{heng18}) corresponds to ${\cal G}_\mp$ (notation of \citealt{malik19}).  Furthermore, what is written as ${\cal L}$ in equation (10) of \cite{malik19} is $C_\star/2F_\star$ in the notation of \cite{heng18}.  The stellar flux at the top of the atmosphere (TOA) is written as $F_{\star,{\rm TOA}}$ in \cite{malik19} and $F_\star$ in \cite{heng18}.  Equation (6) of \cite{malik19} states the flux associated with the stellar beam,
\begin{equation}
F_{{\rm beam},i} = - \mu_\star ~F_\star ~e^{\tau_i/\mu_\star},
\label{eq:stellar_beam}
\end{equation}
which is a positive quantity as $\mu_\star<0$.  With this book-keeping of notation, the beam terms associated with $F_{\uparrow i}$ and $F_{\downarrow i-1}$, respectively, are
\begin{equation}
\begin{split}
&\frac{1}{\chi} \left[ \psi {\cal G}_+ F_{{\rm beam},i-1} - \left( \xi {\cal G}_- + \chi {\cal G}_+ \right) F_{{\rm beam},i} \right], \\
&\frac{1}{\chi} \left[ \psi {\cal G}_- F_{{\rm beam},i} - \left( \xi {\cal G}_+ + \chi {\cal G}_- \right) F_{{\rm beam},i-1} \right].
\end{split}
\end{equation}
In a slight departure from equation (8) of \cite{malik19}, we absorb the $1/\mu_\star$ coefficient associated with the beam terms into ${\cal G}_\pm$ itself,
\begin{equation}
\begin{split}
{\cal G}_\pm =& \frac{1}{2} \left\{ \frac{\omega_0 \left[ 2E \left( 1 - \omega_0 g_0 \right) + \frac{g_0}{\epsilon_2}\right]}{4E\mu^2_\star\left(E-\omega_0\right)\left(1-\omega_0 g_0 \right) - 1} \left[ \mu_\star \pm \frac{1}{2E \left(1 - \omega_0 g_0 \right)} \right] \right. \\
&\pm \left. \frac{\omega_0 g_0}{2 \epsilon_2 E \left( 1 - \omega_0 g_0 \right) } \right\}.
\end{split}
\label{eq:g_factors}
\end{equation}

\cite{malik19} assumed $\epsilon_2=1/2$, which does not correspond to the Eddington ($\epsilon_2=2/3$) or quadrature ($\epsilon_2=1/\sqrt{3}$) closures for the direct beam; $\epsilon_2$ is undefined for the hemispheric closure \citep{mw80,toon89}.  Physically, the choice of $\epsilon_2=1/2$ implies that an extra fraction $\mu_\star g_0$ of the stellar beam is scattered into the forward (downward) direction compared to the backward (upward) direction.  In \texttt{THOR+HELIOS}, we allow $\epsilon_2$ to be a user-specified choice; our chosen default is the Eddington closure ($\epsilon_2=2/3$).

\section{Rayleigh scattering cross sections}
\label{append:rayleigh}

For H$_2$-dominated atmospheres, the Rayleigh scattering cross section is dominated by the contributions of molecular hydrogen and helium.  It generally has the form \citep{su05},
\begin{equation}
\sigma_{\rm gas,scat} = \frac{24 \pi^3}{n_{\rm ref}^2 \lambda^4} \left( \frac{n_r^2 - 1}{n_r^2+2} \right)^2 K_\lambda,
\end{equation}
where $\lambda$ is the wavelength, $n_{\rm ref}$ is a reference number density, $K_\lambda$ is the King factor and $n_r$ is the real part of the index of refraction.

For molecular hydrogen, we have $K_\lambda=1$, $n_{\rm ref} = 2.68678 \times 10^{19}$ cm$^{-3}$ and \citep{cox},
\begin{equation}
n_r = 1.358 \times 10^{-4} \left[ 1 + 7.52 \times 10^{-3} ~\lambda^{\prime-2} \right] + 1,
\end{equation}
where $\lambda^\prime \equiv \lambda / 1$ $\mu$m.  

For helium (He), we have $K_\lambda=1$, $n_{\rm ref} = 2.546899 \times 10^{19}$ cm$^{-3}$ and \citep{su05,thalman14}
\begin{equation}
n_r = 10^{-8} \left[ 2283 + \frac{1.8102 \times 10^{13}}{1.5342 \times 10^{10} - ~\lambda^{\prime-2}} \right] + 1.
\end{equation}

\section{Global conservation of quantities}
\label{append:conservation}

Figure \ref{fig:global_quantities} tracks the evolution of the global mass, angular momentum, energy and radiative fluxes for each of the 4 GCMs presented in the current study. The relative error in total mass of the atmosphere is small: less than $10^{-11}$ in all cases except the isothermal layers case, for which it is $\sim 2\times 10^{-11}$. The poor convergence properties of the radiative transfer in this case apparently compounds upon the errors in the dynamical core. Axial angular momentum (AAM) is not as well conserved as a result of the use of linear momentum equations in the dynamical core \citep{mendonca18b,deitrick20}. However, the errors in AAM plateau as the flow reaches steady state, thus providing a useful convergence metric \citep{read1986}. All three non-isothermal layer simulations reach a steady-state of the flow in $\sim 1000-2000$ days. For the isothermal simulation, we again note the poor convergence, which in this case applies to the flow. The energy (lower left panel) is not necessarily conserved, because of the external forcing. Ideally, this will be conserved once radiative balance is achieved (lower right panel). All simulations except the isothermal layers case have converged, though there is a gap between in the incoming radiation and the end-state outgoing radiation---an error of $\sim 3-7 \%$. This error is dominated by the numerical diffusion processes (hyper-diffusion and sponge layer). The numerical diffusion also causes the slow drift in the energy (lower left panel) in the non-isothermal cases. 

\section{Comparison to Mendonca et al. 2018}
\label{append:comp2mendonca}

\begin{figure}
\begin{center}
\vspace{-0.1in}
\includegraphics[width=\columnwidth]{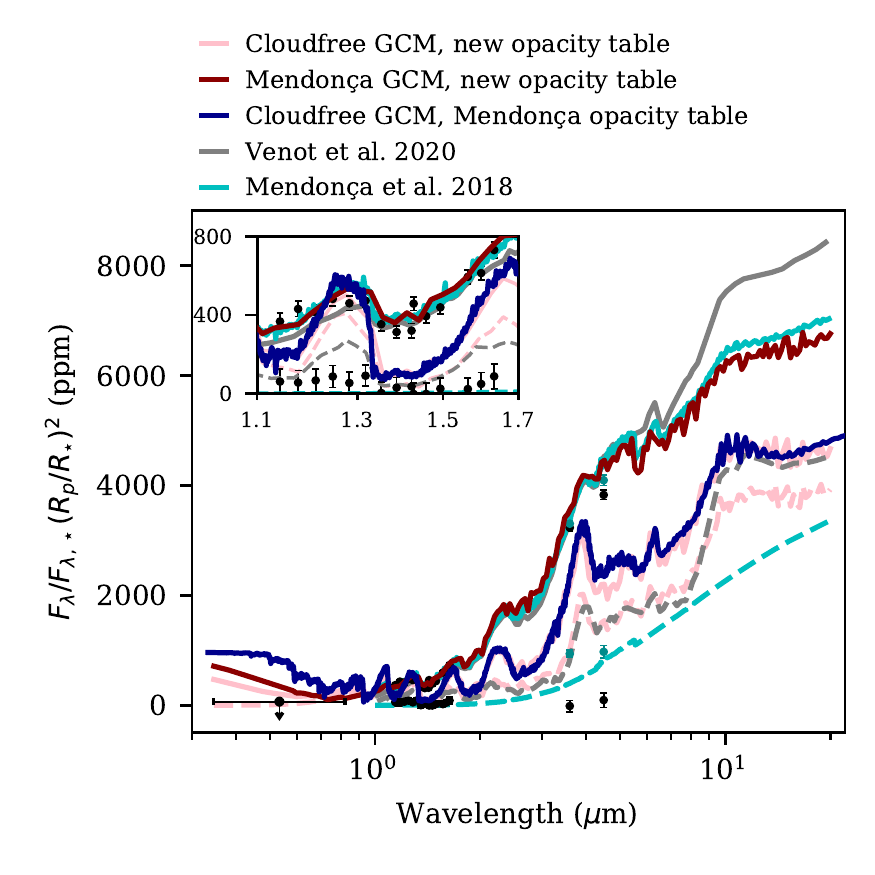}
\end{center}
\vspace{-0.1in}
\caption{\tempbf{Phase dependent spectra from our cloud-free GCM and comparison GCMs \protect\citep{mendonca18a, venot20}. Solid curves are the day-side emission and dashed are the night-side emission. These are identical to the data in Figure \ref{fig:phase_curves}. We have additionally plotted results from 1-D \texttt{HELIOS}. The dark red curve utilized our current opacity table with temperature-pressure data from the day-side of the \protect\cite{mendonca18a} GCM; the dark blue curve utilized the opacity table from \protect\cite{mendonca18a} and temperature-pressure data from the day-side of our current cloud-free GCM.}}  
\vspace{-0.1in}
\label{fig:phase_spec_comp}
\end{figure}

Figure \ref{fig:phase_spec_comp} shows the phase dependent spectra from our cloud-free GCM, compared to the simulations from \cite{mendonca18a} and \cite{venot20}. The other simulations are removed for easier comparison. We have over-plotted spectra from two additional tests. In the first, we have used the hemispherically-averaged, line-of-sight corrected temperature-pressure profile from the day-side of the \cite{mendonca18a} GCM to produce spectra using our current $R=500$ opacity table and 1-D \texttt{HELIOS}. The result is quite similar to the spectrum produced by \cite{mendonca18a}, which used a different opacity table, though with a similar list of sources. In the second, we used the day-side temperature-pressure profile from our cloud-free GCM, produced using the same averaging process, and the opacity table from \cite{mendonca18a} to produce another spectrum using 1-D \texttt{HELIOS}. This result is now quite similar to our post-processed cloud-free GCM. Together, the two results show that the difference between the \cite{mendonca18a} spectrum and ours is not due to the minor differences in opacity tables (i.e., a few different molecules and different resolutions).

Inspecting the temperature-pressure profiles from our cloud-free GCM and the \cite{mendonca18a} GCM, we note that there is a difference of $\sim 500$ K in the photosphere ($\sim0.01-0.1$ bar) on the day-side. The increased long-wave emission in the \cite{mendonca18a} GCM is thus a result of the higher temperatures in this region. For the moment it is unclear why the new model with $k-$tables is so much cooler than the dual-band gray RT model, but we have noticed that this difference tends hold in our testing for other planets.

\label{lastpage}
\end{document}